\documentclass[
amssymb,preprint,preprintnumbers,nofootinbib,superscriptaddress]{revtex4}
\usepackage{amsmath}
\usepackage{graphicx}
\graphicspath{{Figures/}}
\usepackage{latexsym}
\usepackage{amsfonts}
\usepackage{url,hyperref}
\usepackage{bm}
\usepackage{textcomp}
\usepackage{color}
\usepackage{bbm}
\usepackage{slashed}
\usepackage{caption}
\usepackage{epstopdf}
\usepackage{subcaption}
\usepackage{placeins}
\usepackage{color}
\usepackage[a4paper,left=20mm,right=20mm,top=25mm,bottom=25mm]{geometry}

\begin{document}
\title{Thermodynamics of Black String from R\'enyi entropy in de Rham-Gabadadze-Tolley Massive Gravity Theory}

\author{Peerawat Sriling \footnote{Email: p.rawat08@hotmail.com}}
\affiliation{The Institute for Fundamental Study, Naresuan University, Phitsanulok, 65000, Thailand}
\affiliation{Thailand Center of Excellence in Physics, Ministry of Higher Education, Science, Research and Innovation, 328 Si Ayutthaya Road, Bangkok 10400, Thailand}

\author{Ratchaphat Nakarachinda \footnote{Email: tahpahctar\_net@hotmail.com}}
\affiliation{The Institute for Fundamental Study, Naresuan University, Phitsanulok, 65000, Thailand}

\author{Pitayuth Wongjun \footnote{Email: pitbaa@gmail.com}}
\affiliation{The Institute for Fundamental Study, Naresuan University, Phitsanulok, 65000, Thailand}
\affiliation{Thailand Center of Excellence in Physics, Ministry of Higher Education, Science, Research and Innovation, 328 Si Ayutthaya Road, Bangkok 10400, Thailand}

\begin{abstract}
 The de Rham-Gabadadze-Tolley (dRGT) black string solution is a cylindrically symmetric and static solution of the Einstein field equation with graviton mass term. For the asymptotically de Sitter (dS) solution, it is possible to obtain the black string with two event horizons corresponding to two thermodynamic systems. The R\'enyi entropy is one of the entropic forms which is suitable to deal with nonextensive properties of the black string. In this work, we investigated the possibility to obtain a stable black string by using the R\'enyi entropy in both separated and effective approaches. We found that the nonextensivity provides the thermodynamically stable black string with moderate size in both approaches. The transition from the hot gas phase to the moderate-sized stable black string in the separated/effective description is a first-order/zeroth-order phase transition. The significant ways to distinguish the black string from both approaches are discussed.

\end{abstract}
\maketitle{}

\section{Introduction}\label{sect: intro}
One of the important theoretical predictions of General Relativity (GR) is the possibility of the existence of a mysterious object, namely, black hole. Until now, there are several observations confirming the existence of the black holes in nature \cite{Akiyama:2019cqa,Garofalo:2020ajg,Vincent:2020dij,Dokuchaev:2020rye,TheLIGOScientific:2016src}. The simple black hole solutions are based on the spherical symmetry and static condition. For cylindrically symmetric solution, the theoretical predictions of such the similar mysterious object are not frequently found in literature. It may be a consequence from the t' Hooft's conjecture which states that horizons form when and only when a mass gets compacted into a region whose circumference is less than $4\pi GM$ in all directions \cite{Thorne}. According to this conjecture, the cylindrical matter will not form as black hole does. However, the theoretical investigation has been performed in order to overcome the mentioned conjecture. It is found that  the cylindrical solutions are shown to exist in the GR with the existence of the cosmological constant \cite{Lemos:1994xp,Lemos:1994fn,Cai:1996eg,Lemos:1995cm}. With the cylindrical symmetry, the horizons usually is a circular and then such corresponding object is commonly known as the black string. This is one of theoretical evidences suggesting that the t' Hooft's conjecture may be violated in asymptotically de Sitter (dS)/anti-de Sitter (AdS) spacetime. 

There have been many attempts to modify GR due to a cosmological aspect as the universe expands with an acceleration. In this context, many solutions with asymptotically dS/AdS spacetime have been explored and one of those interesting solutions is the de Rham-Gabadadze-Tolley (dRGT) massive gravity solution \cite{deRham:2010ik,deRham:2010kj}. The dRGT massive gravity theory is one of the modified gravity theories in which the suitable mass terms are added into GR. Spherically symmetric solutions in dRGT massive gravity are also found \cite{Berezhiani:2011mt,Brito:2013xaa,Volkov:2013roa,Cai:2012db,Babichev:2014fka,Babichev:2015xha,Hu:2016hpm}. The black hole solutions have been intensively investigated, for example, thermodynamic properties \cite{Cai:2014znn,Ghosh:2015cva,Adams:2014vza,Xu:2015rfa,Capela:2011mh,Hu:2016mym,Zou:2016sab,Hendi:2017arn,Hendi:2017bys,EslamPanah:2016pgc,Hendi:2016hbe,Hendi:2016uni,Hendi:2016yof,Arraut:2014uza,Arraut:2014iba}, greybody factor \cite{Boonserm:2017qcq,Kanzi:2020cyv,Hou:2020yni,Boonserm:2021owk}, quasinormal modes  \cite{Burikham:2017gdm,Wongjun:2019ydo,Burikham:2020dfi}, critical heat engine \cite{Yerra:2020bfx} and phase transition in Ruppeiner geometry \cite{Yerra:2020oph,Yerra:2021hnh}. Since the dRGT solution is one of solutions with asymptotically dS/AdS spacetime, it is possible to obtain the black string solution \cite{Tannukij:2017jtn}. Consequently, the solution for the charged black string as well as rotating black string are explored \cite{Ghosh:2019eoo,Hendi:2020apr}. 

One of important clues in linking general relativity, thermodynamics, and quantum field theory is a discovery that a black hole behaves as a thermal object \cite{Hawking:1974sw,Bekenstein:1973ur,Bardeen:1973gs}. Actually, the discovery is rooted from the idea that the black hole can emit the radiation and then its area can be thought as the entropy. This idea cannot be applied only for the black hole but also for the black string \cite{Lemos:1994fn,Cai:1996eg}. Moreover, the greybody factor from the black string has been investigated \cite{Ahmed:2017edq}. In the context of dRGT massive gravity, the thermodynamic properties \cite{Tannukij:2017jtn,Ghosh:2019eoo,Hendi:2020apr} and greybody factor \cite{Boonserm:2019mon} of the dRGT black string have been investigated. 

It is worthwhile to note that the entropy of the black hole/string is relevant in the Planck scale and the microscopic degrees of freedom should be counted by using proper quantum gravity which is still debatable and not completely successful to construct. Therefore, any classical gravity theories including GR, dRGT massive gravity and other classical modified gravity theories are not supposed to be used to describe the microscopic nature on the entropy of the black hole/string. In particular, for the dRGT massive gravity, there exists the cutoff scale $r_{\Lambda_3}$ which is much larger than the Planck length and then the theory does not rely on a viable regime to investigate the thermodynamics in the aspect of the microscopic description. However, it may be possible to investigate the thermodynamical properties of the black hole/string in the context of macroscopic description as found in literature. For example, the equivalent thermodynamical equations can be obtained by using the Einstein field equations evaluated at the horizons \cite{Padmanabhan:2002sha,Paranjape:2006ca,Kothawala:2007em}. In this paper, we investigate the thermodynamical properties by restricting our consideration on macroscopic description. In fact, the dRGT massive gravity can be recovered GR at a scale below the Vainshtein radius $r_V$ which is much larger than the cutoff scale. Therefore, the thermodynamical properties affected by the graviton mass is supposed to be modified at the scale comparable to the Vainshtein radius. In particular, we will investigate this effect and this is one of the main aims of the present paper.

It is important to note that the investigation on the black string is mostly performed in the asymptotically AdS spacetime. This comes from the fact that dS black string has no horizon, then it does not  correspond to thermodynamic system. For the dRGT black string with asymptotically dS spacetime, even though there exist horizons, the corresponding thermodynamic systems are found to be unstable. Moreover, due to the existence of two horizons, the black string corresponds to two systems at which their temperatures are generically different and then they are out of thermal equilibrium. Therefore, thermodynamic studies for this kind of black string have encountered a number of difficulties in considerations. This issue is also found in the black hole case. 
For the multi-horizon issue, the investigation can be performed in two different approaches; separated system \cite{Kubiznak:2015bya} and effective system \cite{Urano:2009xn,Ma:2013aqa,Zhao:2014raa,Zhang:2014jfa,Ma:2014hna,Guo:2015waa,Guo:2016eie,Simovic:2020dke} approaches. For separated system approach, we can treat the systems to be in quasiequilibrium state. The thermodynamic systems can be investigated separately by assuming that the systems are separated far enough and the temperatures of the systems are not significantly different. For the effective system approach, one can treat the systems as a single system describing by the effective thermodynamic quantities. For the stability issue, it is possible to obtain the stable black hole by considering the R\'enyi entropy instead of the usual Gibbs-Boltzmann (GB) entropy \cite{Czinner:2015eyk,Czinner:2017tjq,Tannukij:2020njz,Promsiri:2020jga,Promsiri:2021hhv,Nakarachinda:2021jxd}. This may be a consequence from the fact that the entropy of the black hole is proportional to its area and then corresponds to the nonextensive thermodynamic system. 

The entropy plays an essential role in the GB statistics. It is a macroscopic thermodynamic variable which is directly analogous to the entropic function of a random variable in information theory, known as the Shannon entropy. Actually, the entropy can be viewed an average value of a surprisal, $S =- \sum_i P_i \log P_i$ where the surprisal is defined as $\log(1/ P_i)$ and $P_i$ is a probability to find the state $i^{th}$. In the context of thermodynamics, the equilibrium state can be properly defined by requiring the maximum entropy. Consequently, the empirical temperature of the equilibrium state can be defined via the zeroth law as $1/T = \partial S/ \partial E$ where $E$ is the energy of the system. One of crucial properties of the entropy is that it is an extensive quantity which satisfies $S \propto N \propto V$  where $N$ and $V$ are number of particles and volume of the system, respectively. 

Even though the GB statistics is widely used for many aspects of thermodynamic system, it is still not enough to explain a system with long-range interaction, e.g., astronomical objects \cite{Jiulin:2004bg}. For this situation, the system is considered as the nonextensive one and then the entropy should be generalized to cover the nonextensivity. One of the useful generalized entropies is known as the Tsallis entropy \cite{Tsallis:1987eu}. The nonextensivity of the Tsallis entropy can be seen from the nonadditive composition rule as $S^{\text{12}}_T = S^\text{1}_T + S^\text{2}_T + \lambda S^\text{1}_T S^\text{2}_T$ where $S^{\text{12}}_T$ is the Tsallis entropy of the entire system, $S^{\text{1}}_T$ and $S^{\text{2}}_T$ are the Tsallis entropies of the two separated systems, and $\lambda$ is the nonextensive parameter. Even though the Tsallis entropy may be compatible to the nonextensive system with long-range interaction, it is not suitable to be used to define the empirical temperature of equilibrium state as found in the GB case. This is due to the nonadditive composition nature of the Tsallis entropy itself. By restricting to the nonextensive system, there is the formal logarithm map from the Tsallis entropy to one which obeys additive composition rule \cite{Biro:2011} $S_R = \log(1 + \lambda S_T)/\lambda$, known as the R\'enyi entropy \cite{Renyi:1959,Renyi:1961}. 

Since the black hole/string entropy is proportional to the area instead of the volume, the suitable form of the black hole/string entropy can be thought as the Tsallis entropy. However, in order to properly define the empirical temperature satisfying the zeroth law, the R\'enyi entropy is worthy choice while the nonextensive nature of the black hole/string is still relevant. It has been found that the nonextensive parameter can play the crucial role to the stability and phase transition of the dS black holes  \cite{Tannukij:2020njz,Nakarachinda:2021jxd}. In the present work, we apply this strategy to the dRGT black string by including nonextensive effect from the R\'enyi entropy. For the separated system approach, we analyze the stability of black string by considering the temperature profile and heat capacity. We found that the nonextensivity significantly provides the local stability of the  black string. The lower bound of the nonextensive parameter is obtained. By analyzing the Gibbs free energy, we found that it is possible to obtain the globally stable black string with stronger bound on the nonextensive parameter. We also found that it is possible to obtain the first-order Hawking-Page phase transition which is the transition between the thermal radiation or hot gas phase and the stable black hole phase. For the effective system approach, we use the suitable definition of the effective quantities in Ref. \cite{Nakarachinda:2021jxd} to perform the stability analysis in the same way as done in separated system approach. The bound on the extensive parameter is obtained and the significant ways to distinguish the black string from both approaches are discussed. The Hawking-Page phase transition for the effective system approach is analyzed and found that it is the zeroth-order. This is one of significant results which is different from one in separated system approach.

This paper is organized as follows. In Sec. \ref{sect:dRGT BS}, we review the dRGT black string solution and then analyze horizon structure. In Sec. \ref{sect: Separated thermo} we investigate thermodynamic properties of two separated systems using the R\'enyi entropy. In Sec. \ref{sect: Effective thermo}, the entire thermodynamic system consisting of two horizons with different temperatures is considered in such a way that it is a single effective system. Finally, we conclude with remarks of the effect of nonextensivity and the validity of our approach in giving a clear physical implication in Sec. \ref{sec: concl}.


\section{dRGT black string}\label{sect:dRGT BS} 

In this section, the dRGT massive gravity is reviewed. One of the black string solutions, which is significantly different to that obtained in GR with cosmological constant. We are also interested in the dS branch of the solution. The structure of the horizons will be discussed.

\subsection{dRGT massive gravity}
The four-dimensional action of the well-known dRGT massive gravity theory is given by \cite{deRham:2010ik,deRham:2010kj}
\begin{eqnarray}
	S=\int d^4x\sqrt{-g}\frac{M^2_{P}}{2}\left[R+m^2_{g}\,\mathcal{U}(g,f)\right], \label{SdRGT}
\end{eqnarray}
where $R$ and $M_P$ are the Ricci scalar and the Planck mass, respectively. The interaction term consists of the graviton mass parameter $m_g$ and the non-trivial potential $\mathcal{U}$ expressed as
\begin{eqnarray}
	\mathcal{U}&=&\mathcal{U}_2+\alpha_3\mathcal{U}_3+\alpha_4\mathcal{U}_4,\\
	\mathcal{U}_2&=&[\mathcal{K}]^2-[\mathcal{K}^2],\\
	\mathcal{U}_3&=&[\mathcal{K}]^3-3[\mathcal{K}][\mathcal{\mathcal{K}}^2]+2[\mathcal{K}^3],\\
	\mathcal{U}_4&=&[\mathcal{K}]^4-6[\mathcal{K}]^2[\mathcal{K}^2]+8[\mathcal{K}][\mathcal{K}^3]+3[\mathcal{K}^2]^2-6[\mathcal{K}^4],
\end{eqnarray}
where $\alpha_3$ and $\alpha_4$ are free parameters of the theory. The above potential is indeed a suitable form in which the ghost instability is eliminated. The quantity $[\mathcal{K}^n]$ denotes the trace of the $n$-th power of the matrix
\begin{align}
	\mathcal{K}^{\mu}_{\hspace{.2cm}\nu}=\delta^{\mu}_{\nu}-\big(\sqrt{g^{-1}f\,}\,\big)^{\mu}_{\hspace{.2cm}\nu}.
\end{align}
It is noticed that this ghost-free theory is not constructed only from the physical metric $g_{\mu\nu}$ but also the reference or fiducial metric $f_{\mu\nu}$. The fiducial metric $f_{\mu\nu}$ is non-dynamical field. Therefore, the field equations with respect to $f_{\mu\nu}$ is just the constraints. In other words, $f_{\mu\nu}$ plays a role of the Lagrange multiplier in order to construct the suitable form of the mass terms.

By varying the action in Eq. \eqref{SdRGT} with respect to the physical metric $g_{\mu\nu}$, the modified Einstein equations are obtained as follows:
\begin{align}
	G_{\mu \nu}+m_{g}^2X_{\mu\nu}=0, \label{EOM}
\end{align}
where $G_{\mu \nu}$ is the Einstein tensor and $X_{\mu\nu}$ is the effective energy-momentum tensor associated with a part from varying the potential term. This effective energy-momentum tensor can be written in the explicit form as
\begin{align}
	X_{\mu\nu}=\mathcal{K}_{\mu\nu}-[\mathcal{K}]g_{\mu\nu}-\alpha\Big(\mathcal{K}_{\mu\nu}^2-[\mathcal{K}]\mathcal{K}_{\mu\nu}+\frac{\mathcal{U}_2}{2}g_{\mu\nu}\Big)+3\beta\Big(\mathcal{K}_{\mu\nu}^3-[\mathcal{K}]\mathcal{K}_{\mu\nu}^2+\frac{\mathcal{U}_2}{2}\mathcal{K}_{\mu\nu}-\frac{\mathcal{U}_3}{6}g_{\mu\nu}\Big). \label{X}
\end{align}
Here, the free parameters $\alpha_3$ and $\alpha_4$ are redefined in convenient form as
\begin{align}
	\alpha_3=\frac{\alpha-1}{3},\hspace{1cm} 
	\alpha_4=\frac{\beta}{4}+\frac{1-\alpha}{12}.
\end{align}
According to the the Bianchi identity of the Einstein tensor, the effective energy-momentum tensor $X_{\mu\nu}$ is covariantly divergent free
\begin{align}
	\nabla^{\mu}X_{\mu\nu}=0, \label{Con}
\end{align}
where $\nabla^{\mu}$ denotes the covariant derivative associated with the metric $g_{\mu\nu}$. Note that these constraints can also be derived from varying the action with respect to the fiducial metric.

\subsection{dRGT black string solution and horizon structure}
In this subsection, we are interested in one of the cylindrically symmetric solutions in dRGT massive gravity which is significantly different to the solution in GR with the cosmological constant. The line element for this solution can be written as
\begin{eqnarray}
	ds^2=-\mathcal{F}(r)dt^2+\frac{dr^2}{\mathcal{F}(r)}+r^2\left(d\varphi^2+\alpha^2_gdz^2\right),\label{gen metric}
\end{eqnarray}
where $\alpha_g$ is a constant in the unit of mass. Let us choose the fiducial metric in the form of \begin{eqnarray}
	f_{\mu\nu}=\text{diag}\Big(0, 0, h^2, h^2\alpha_g^2\Big),
\end{eqnarray}	
where the constant $h$ plays the role of the radial function similar to the radial coordinate $r$ in the physical metric. Substituting the above ansatz metrics to the field equation \eqref{EOM}, the function $\mathcal{F}(r)$ in the line element in Eq. \eqref{gen metric} is eventually given by \cite{Tannukij:2017jtn}
\begin{eqnarray}
	\mathcal{F}(r)=-\frac{4M}{r}-m_g^2\left(r^2-c_1r-c_0\right),
\end{eqnarray}
where $M=M_\text{ADM}/\alpha_g$ and $M_\text{ADM}$ is the Arnowitt-Deser-Misner mass per unit length of the $z$-coordinate. Emphasize that the mass parameter $M$ is indeed in the unit of length, since the Newtonian gravitational constant $G$ is set to be unity in this consideration. The model parameters $c_1$ and $c_0$ are respectively expressed in terms of the aforementioned parameters as follows:
\begin{eqnarray}
	c_1\equiv-h\left(\frac{1+2\alpha+3\beta}{1+\alpha+\beta}\right),\quad
	c_0\equiv h^2\left(\frac{\alpha+3\beta}{1+\alpha+\beta}\right).
\end{eqnarray}

\begin{figure}[h!]
	\begin{center}
		\includegraphics[scale=0.42]{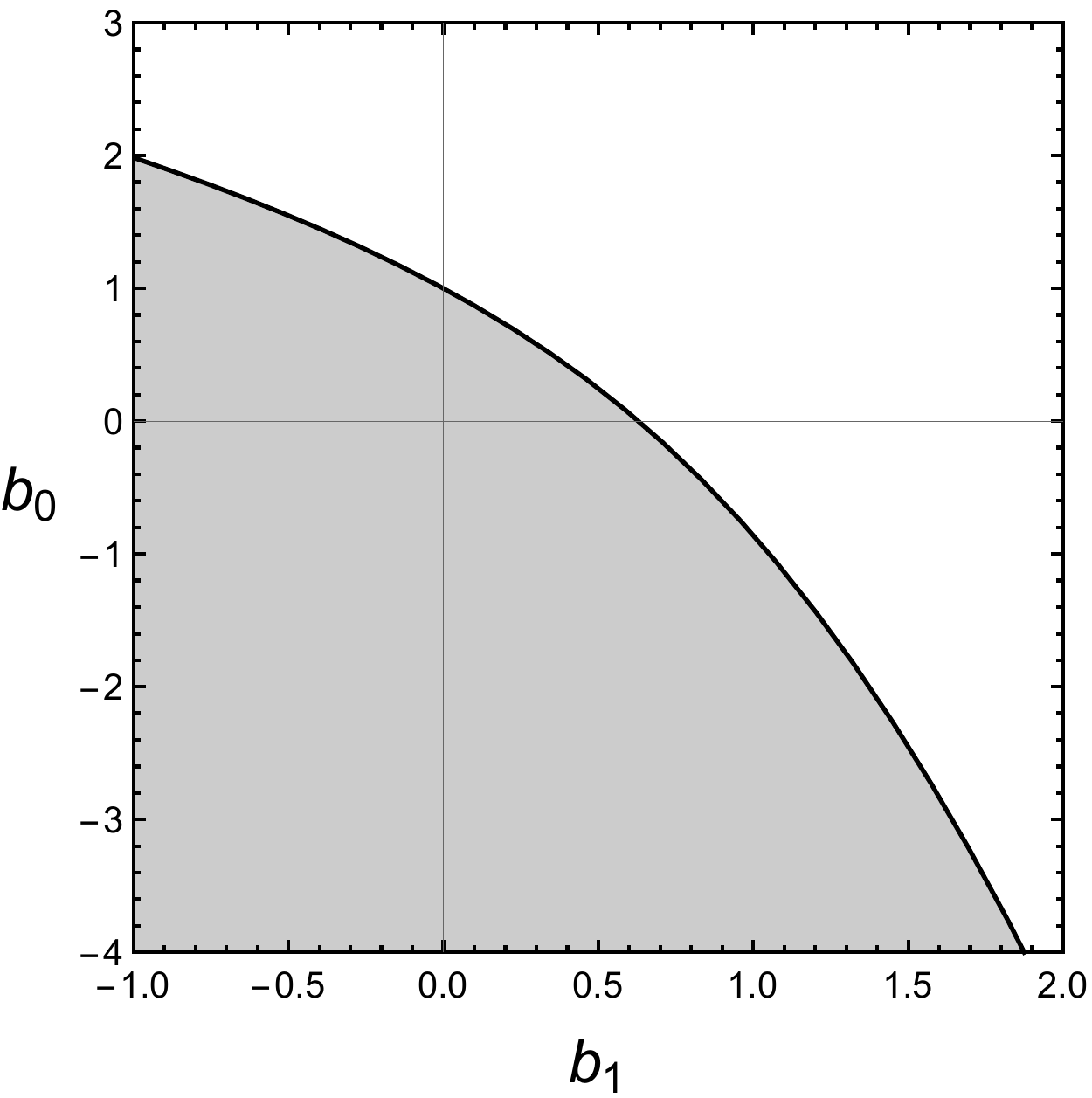}\hspace{0.2cm}
		\includegraphics[scale=0.4]{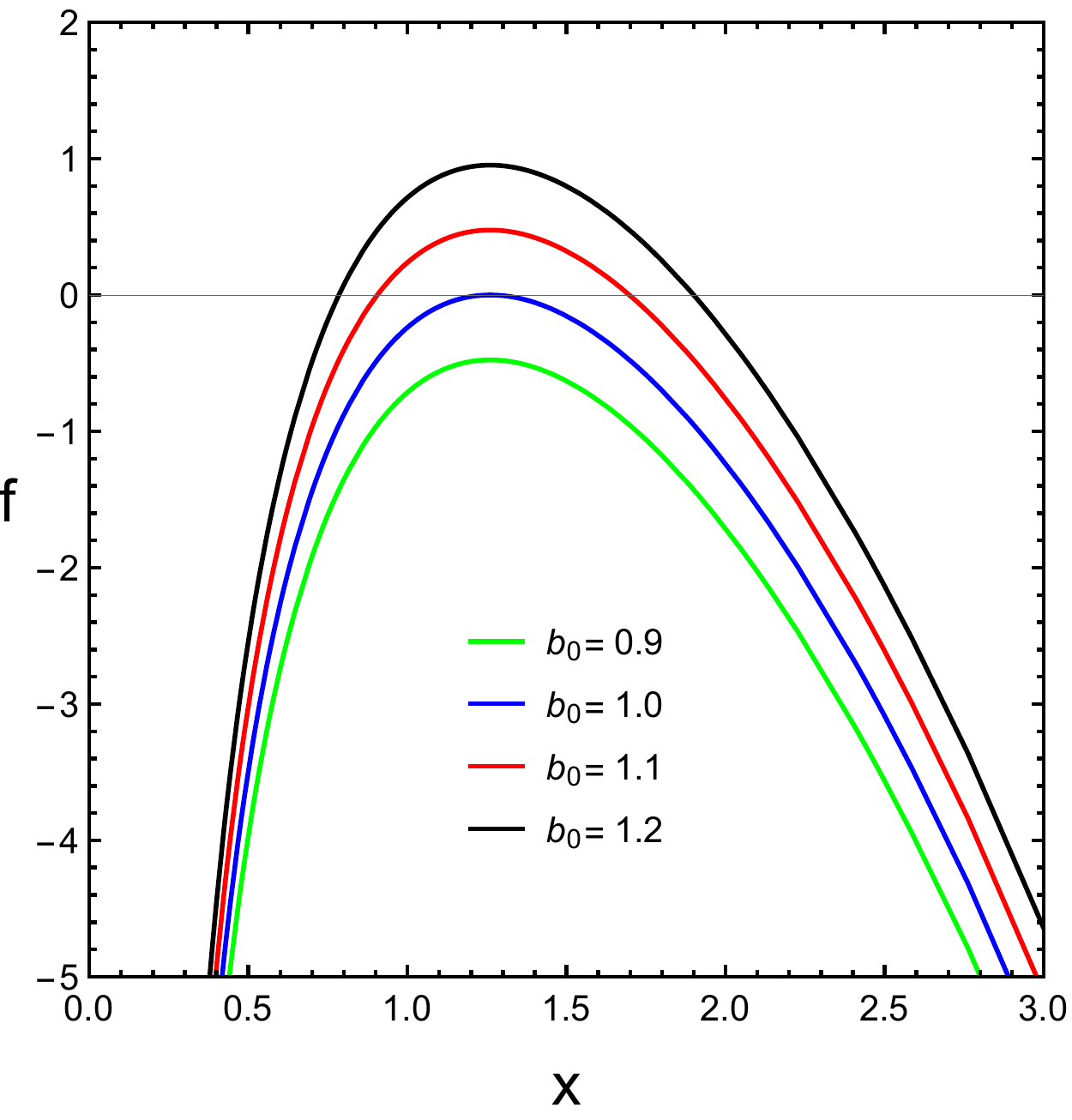}\hspace{0.2cm}
            \includegraphics[scale=0.4]{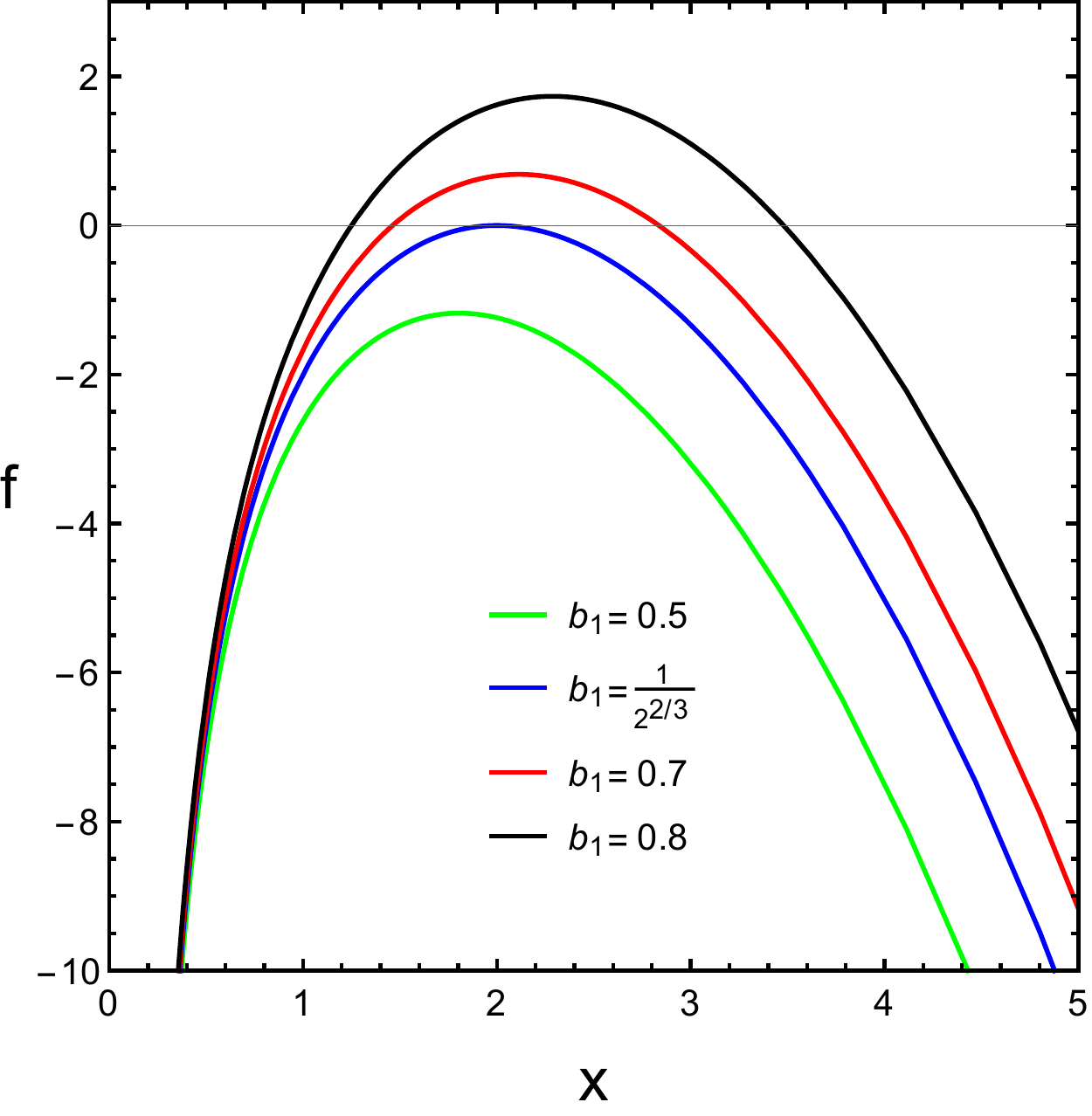}
	\end{center}
	\caption{The left panel shows the region of the existence of the horizons in $(b_1,b_0)$ space. The middle and right panels show the behaviors of the function $f(x)$ for various values of $b_0$ (with fixing $b_1=0$) and $b_1$ (with fixing $b_0 =0$), respectively.}\label{fig:horizon}
\end{figure}

The horizons of the black string can be defined in the same way as found in black hole case by solving $\mathcal{F}(r)=0$. As a result, the number of possible solutions depends on the sign of $m_g^2$. Now, we will investigate the structure of the horizons by restricting our attention on the asymptotically dS solution, i.e., $m_g^2>0$. In the limit $c_0=c_1=0$ and $m_g=\alpha_g$, the dRGT black string solution reduces to the Lemos' black string solution \cite{Lemos:1994xp}
\begin{eqnarray}
	\mathcal{F}(r)=-\frac{4M}{r}-\alpha_g^2r^2.
\end{eqnarray}
For this case, one can see that it is not possible to obtain the horizons because $\mathcal{F}(r)$ is always negative. It implies that the effects of $c_0$ and $c_1$ are significantly required for the existence of the horizons. Therefore, the structure of graviton mass is necessarily important. For convenience, we rewrite the function $\mathcal{F}(r)$ in terms of dimensionless quantities by redefining parameters as follows:
\begin{eqnarray}
    r=xr_V,\quad
    c_1=3\times2^{2/3}b_1r_V,\quad
    c_0=3\times2^{2/3}b_0r_V^2,\quad
    r_V=\left(\frac{M}{m_g^2}\right)^{1/3}.
\end{eqnarray}
Here, $r_V$ is a physical length scale at which the graviton mass plays a major part of the gravitational interaction for the radius being much larger than the Vainshtien radius $r\gg r_V$, while it is suppressed at the scale below the Vainshtien radius $r\ll r_V$. As a result, the metric function can be rewritten as
\begin{align}
	f(x)= \frac{r_V}{M}\mathcal{F}(r)=-\frac{4}{x}-x^2+\left(3\times2^{2/3}b_1\right)x+\left(3\times2^{2/3}b_0\right).\label{f in x}
\end{align}
It is important to note that the dRGT massive gravity theory is an effective field theory so that there exists the cutoff scale as $r_{\Lambda_3} = (m_g^2 M_\text{P})^{-1/3}$. Below this length scale, the theory is supposed to be not trustable. Therefore, the dRGT black string with the radius smaller than $r_{\Lambda_3}$ is untrustworthy to investigate. However, the radius of the dRGT black string is scaled by the Vainshtien radius $r_V$ which is generically larger than the cutoff scale, $r_V = (m/M_\text{P})^{1/3}r_{\Lambda_3}$ where $m = M/G \sim M_\text{P}^2 M$. For example, by using the black string mass of $100 M_\text{S}$ where $M_\text{S}$ is the Sun's mass, the Vainshtein radius is much larger than the cutoff scale as $r_V \sim 10^{13} r_{\Lambda_3}$. For the graviton mass satisfying the current observation $m_g \sim 10^{-23} \,eV$ \cite{LIGOScientific:2020tif}, the cutoff scale can be estimated as $10$ m.  As a result, the thermodynamic quantities such as the temperature and entropy for the dRGT black string are reliably investigated. This can be seen explicitly in the next sections.   From Eq. \eqref{f in x}, it is seen that the horizon structure now can be charaterized by two parameters $b_0$ and $b_1$. The extremum can be found by solving $f'(x)=0$, where the prime denotes the derivative with respect to its argument, then we obtain 
\begin{align}
	3\times2^{2/3} b_1-\frac{4}{x^2}+2x=0.
\end{align}
Obviously, the parameter $b_0$ (or $c_0$) does not influence the extremum of $f(x)$. Therefore, we obtain the extremum as
\begin{align}
	x_0=\frac{1}{2^{1/3}}\Big(b_1+\frac{b_1^2}{B_1}+B_1\Big), \quad	
	B_1=\left(2+b_1^3+2\sqrt{1+b_1^3}\right)^{1/3}. \label{eq:xex}
\end{align}
Notice that the extremum of $f(x)$ depends only on $b_1$. Substituting the above value to Eq. \eqref{f in x}, the value of the function $f(x)$ at extremum is written as
\begin{align}
	f(x_0)
	=\frac{6}{2^{1/3}} b_0-\frac{1}{2^{2/3}}\left(b_1+\frac{b_1^2}{B_1}+B_1\right)^2+\frac{6 b_1}{2^{2/3}} \left(b_1+\frac{b_1^2}{B_1}+B_1\right)-\frac{4\times2^{1/3}}{\left(b_1+\frac{b_1^2}{B_1}+B_1\right)}.
\end{align}
In order to obtain the condition for existence of the horizons, one requires that $f(x_0) \geq 0$. As a result, one can find the relation between $b_1$ and $b_0$ to satisfy this condition as follows:
\begin{align}
b_0 \geq \frac{\left(b_1+\frac{b_1^2}{B_1}+B_1\right)^3-6b_1 \left(b_1+\frac{b_1^2}{B_1}+B_1\right)^2+8}{6\times2^{1/3} \left(b_1+\frac{b_1^2}{B_1}+B_1\right)}.
\end{align}
The region for which there exist horizons can be illustrated in the left panel in Fig. \ref{fig:horizon}, while the shaded region corresponds to a no-horizon region. From this figure, one can see that it is not possible to have horizon at the origin  $(b_1,b_0) = (0,0)$. Therefore, there is no horizon for usual black string as we have found.  From Eq. \eqref{eq:xex}, one can find condition on $b_1$ as $b_1 > -1$. This is a requirement to have the positive real root of the extremum. It is more convenient for us to consider the case $b_1 \sim 0$, since we can analyze the possibility of finding a deviation from the Lemos' black string. In this case, one has a condition on $b_0$ as $b_0 \gtrsim 1- 2^{1/3} b_1$.

Thermodynamic properties of the black string can be investigated in the same fashion as in the black hole case. The thermodynamic functions of the black string can be obtained by evaluating at the event horizons. For example, the temperature can be obtained by evaluating the quantities $\frac{\mathcal{F}'(r)}{4\pi}$ at the horizons. As a result, the black string or black hole with multiple horizons yields the thermodynamic systems with different temperatures. Therefore, it corresponds to non-equilibrium thermodynamic states. In this case, we need more careful way to deal with this kind of black string. In this work, we separate our consideration into two classes; separated system approach and effective system approach. For the separated system approach, the thermodynamic systems can be investigated separately by assuming that the systems are separated far enough and the temperatures of the systems do not significantly differ. For the effective system approach, one can treat the systems as a single system describing by the effective thermodynamic quantities.


\section{Separated thermodynamic systems}\label{sect: Separated thermo}

In this section, we will investigate the thermodynamic properties of dRGT black string by considering the defined thermodynamic system of each horizon separately. The systems are assumed to be very far away from each other. The black string mass $M$ in terms of horizon radius can be found by solving $\mathcal{F}(r_h)=0$, and then it can be expressed as
\begin{align}
M(r_h)=-\frac{r_h}{4}m_g^2(r_h^2-c_1r_h-c_0). \label{M-horizon}
\end{align}
The Hawking temperature of the dRGT black string can be identified via the surface gravity at the horizon as $T(r_h)=\frac{|\kappa|}{2\pi}=\frac{|\mathcal{F}'(r_h)|}{4\pi}$. As a result, temperature can be written in terms of the horizons as 
\begin{align}
T_{b,c}&=\pm\frac{m_g^2(c_0+2c_1 r_{b,c} - 3 r^2_{b,c})}{4\pi r_{b,c}}. \label{eq:T-kappa}
\end{align}
Note that, in our consideration, we restrict on the dS type of the solution. Therefore, there are only two horizons: the smaller one denotes by black string horizon, $r_b$ and the larger one denotes by the cosmic horizon, $r_c$. Here, the plus and minus signs in Eq. \eqref{eq:T-kappa} denote the temperatures for black string horizon and cosmic horizon, respectively. Moreover, the quantities with subscripts $``b"$ and $``c"$ are  the quantities defined for the system of black string horizon and cosmic horizon, respectively. As a result, the dRGT black string corresponds to two thermodynamic systems with different temperatures. Note also that both temperatures are positive for whole viable range of parameters.

The entropy of the systems can be defined in the same manner as found in the black hole case by using the area law. As a result, the entropy known as the Bekenstein-Hawking entropy can be expressed as
\begin{align}
S_{BH}=\frac{\pi r_h^2}{2},\label{S GB}
\end{align}
where $S_{BH}=A/(4\alpha_g)$ and $A=2\pi r_h\times\alpha_g r_h$ is the area of the cylinder shell per unit length of the $z$-coordinate. Hence, the above entropy $S_{BH}$ is actually in the unit of the square of length. Note that the definitions of the thermodynamic quantities defined in this way satisfy the first law of thermodynamics, $dM =\pm  T_{b,c} dS_{BH}$. However, the first law of thermodynamics can be further extended, since there are more parameters in the theory which can be promoted to be the thermodynamic variables. In order to further study the thermodynamics of the black string with more general form of the first law, let us consider the Smarr formula of the black string by treating the mass parameter $M$ in Eq. \eqref{M-horizon} as a homogeneous function of all parameters in the theory. As a result, we found that $M$ can be written as the homogeneous function with degree $1/2$ as $M=M(S_{BH},m_g^{-2}, c_1^2, c_0)$. By using Euler's theorem, the black string mass can be written as
\begin{align}
	\frac{1}{2}M &= S_{BH} \frac{\partial M}{\partial S_{BH}}+ m_g^{-2} \frac{\partial M}{\partial m_g^{-2} } + c_1^2 \frac{\partial M}{\partial c_1^2 } + c_0 \frac{\partial M}{\partial c_0}.\label{Euler thm}
\end{align}
We choose to define the thermodynamic pressure proportional to $m_g^2$ given by
\begin{eqnarray}
	P=\frac{3}{8\pi}m_g^2.
\end{eqnarray}
The conjugate variables of $S_{BH}, P, c_1$ and $c_0$ can be respectively computed as follows: 
\begin{eqnarray}
	T &=& \pm \frac{\partial M}{\partial S_{BH}}=T_{b,c},\label{T Smarr}\\
	V&=& \frac{\partial M}{\partial P }=\frac{2 \pi  r_h^3}{3} \left(-1+\frac{c_1}{r_h}+\frac{c_0}{r_h^2}\right),\label{Vmg}\\  
	\Phi_1&=&\frac{\partial M}{\partial c_1}=\frac{2}{3}\pi Pr_h^2,\\  
	\Phi_0&=&\frac{\partial M}{\partial c_0}=\frac{2}{3}\pi Pr_h.
\end{eqnarray}
Therefore, the Smarr formula can be rewritten as 
\begin{align}
	M &=\pm 2TS_{BH}-2PV+\Phi_1 c_1 + 2\Phi_0 c_0.\label{M-Smarr}
\end{align}
It is important to note that the temperature defined in Eq. \eqref{T Smarr} is exactly the same with one defined via the surface gravity in Eq. \eqref{eq:T-kappa}. Moreover, the thermodynamic volume in Eq. \eqref{Vmg} is the volume of the black string (the term without $c_1$ and $c_0$) including the structure of the graviton mass ($c_1$ and $c_0$ terms). For the volume $V$, this expression is already absorbed $\alpha_g$ in the same fashion as $M$ and $S$ in Eqs. \eqref{M-horizon} and \eqref{S GB}, respectively. Therefore, its unit is the cubic of the length not the volume per unit length in $z$-coordinate. Interestingly, it is possible to obtain the positive thermodynamic volume and pressure for the solution in the dRGT massive gravity while, in the solution in GR with cosmological constant, either volume or pressure should be negative as follows:
\begin{eqnarray}
	P_{GR}=\pm\frac{3}{8\pi}m_g^2,\quad
	V_{GR}=\mp\frac{2 \pi r_h^3}{3}.
\end{eqnarray}
Here, all parameters are chosen to have only positive values. Therefore, the structure of the graviton mass can be treated as corrections to the thermodynamic volume. 
As a result, the first law of thermodynamics corresponding to the dRGT black string can be written as 
\begin{align}
	dM = \pm T dS_{BH} +VdP + \Phi_1 dc_1 + \Phi_0 dc_0.
\end{align}
If one considers the system in which the pressure, $c_1$ and $c_0$ are held fix, the system will reduce to the standard thermodynamic system with the first law $dM = \pm TdS_{BH}$. In our work, we will focus on the thermodynamic system with fixing $c_1$ and $c_0$. Therefore, the first laws of the thermodynamic system evaluated at the black string horizon and cosmic horizon can be written as  
\begin{eqnarray}
	dM= \pm T_{b,c} dS_{b,c} +V_{b,c}dP,
\end{eqnarray}
where
\begin{eqnarray}
	S_{b,c}=S_{BH}|_{r_h=r_{b,c}},\quad
	V_{b,c}=V|_{r_h=r_{b,c}}.
\end{eqnarray}

Usually, the entropy of a thermodynamic system is an extensive variable which is proportional to the volume or the number of particles of the system. However, as we have seen, the black string entropy is proportional to the surface area of the black string's horizons, instead of its volume. Therefore, the black string entropy is not the extensive variable. In this context, it is worthwhile to investigate the thermodynamic system by using the nonextensive entropy. In this work, we will use the  R\'{e}nyi entropy which can be written in terms of the Bekenstein-Hawking entropy as \cite{Czinner:2015eyk}
\begin{align}
	S_R=\frac{1}{\lambda}\ln(1+\lambda S_{BH}),
\end{align}
where $\lambda$ is the nonextensive parameter and valid the range $-\infty<\lambda<1$. Note that, in the limit $\lambda\rightarrow0$, the R\'{e}nyi entropy reduces to the Bekenstein-Hawking entropy. Summarily, the first law of thermodynamics can be written as 
\begin{align}
	dM=T_{R(b)}dS_{R(b)}+V_bdP,\quad
	dM=-T_{R(c)}dS_{R(c)}+V_cdP. \label{1stlaw-Renyi}
\end{align}
In this work, we give the pressures of these systems are defined as $P=\frac{3}{8\pi}m_g^2$, and apply the first law of thermodynamics by using the R\'{e}nyi entropy instead of $S_{BH}$. Obviously, the thermodynamic volumes $V_{b,c}$ are not modified due to the nonextensivity effect, since both $M$ and $P$ are independent of the nonextensive parameter $\lambda$. As a result, the R\'{e}nyi temperature of both systems can be written as
\begin{eqnarray}
	T_{R(b)}&=&\left(\frac{\partial M}{\partial S_{R(b)}}\right)_P 
	=\frac{m_g^2(c_0+2c_1 r_b - 3 r^2_b)}{4\pi r_b}\left(1+\lambda \frac{\pi r_b^2}{2}\right)
	=T_b\left(1+\lambda \frac{\pi r_b^2}{2}\right),\\
	T_{R(c)}&=&-\left(\frac{\partial M}{\partial S_{R(c)}}\right)_P 
	=\frac{-m_g^2(c_0+2c_1 r_c - 3 r^2_c)}{4\pi r_c}\left(1+\lambda \frac{\pi r_c^2}{2}\right)
	=T_c\left(1+\lambda \frac{\pi r_c^2}{2}\right).
\end{eqnarray} 
The pressure defined in this way can be written in terms of the temperature as
\begin{align}
	P = \frac{\pm3 r_{b,c} T_{R(b,c)}}{ \left(\pi  \lambda  r_{b,c}^2+2\right) \left(c_0+2c_1r_{b,c}-3r_{b,c}^2\right)}.
\end{align} 
One can see that the pressure is directly proportional to the temperature similar to the case of ideal gas. In the thermodynamic view point, this is not possible to have critical phenomena. In order to obtain such a phenomenon, one may further consider the charged black string.

\begin{figure}[h!]
	\begin{center}
		\includegraphics[scale=0.5]{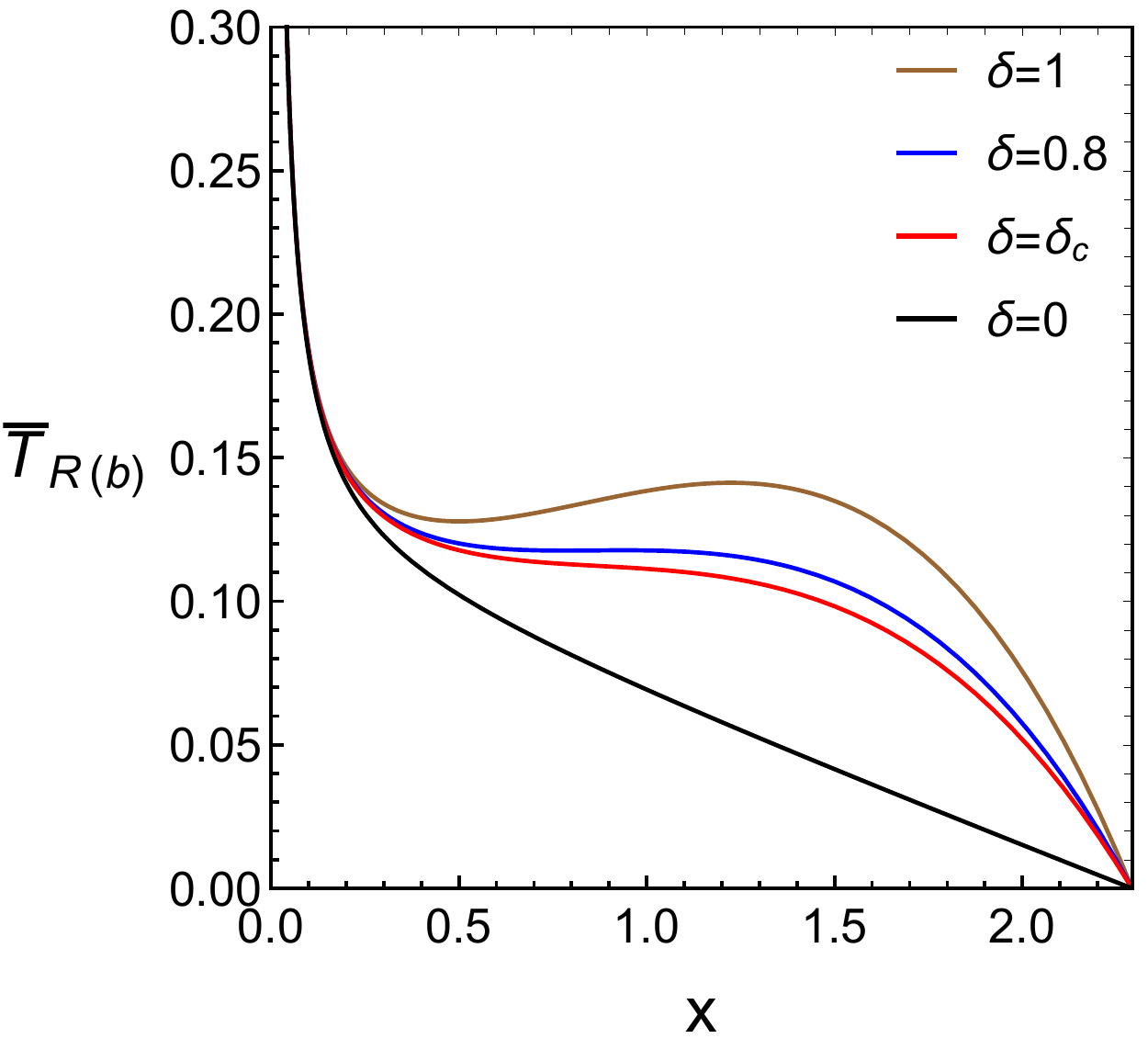}\quad
             \includegraphics[scale=0.5]{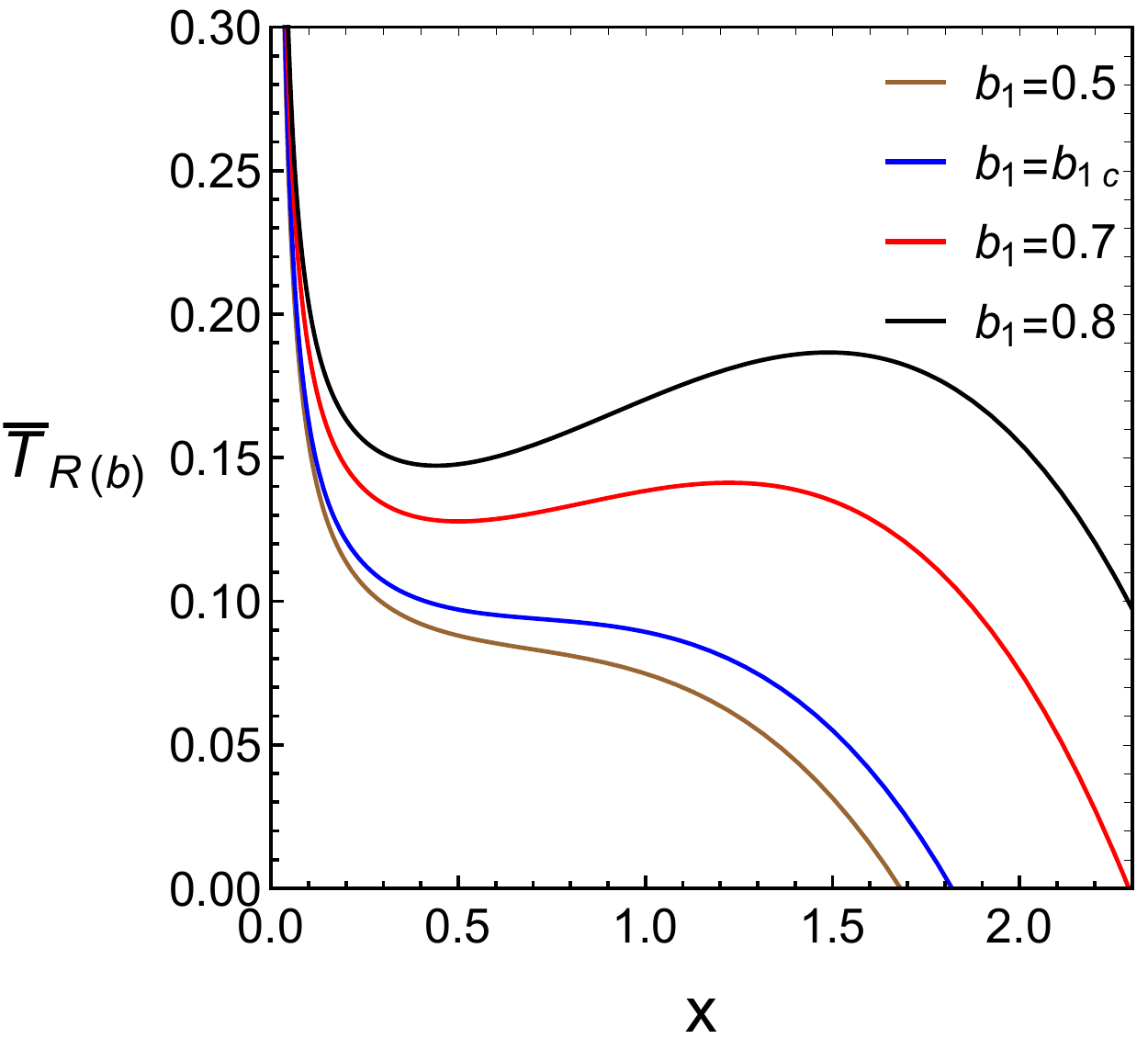}
	\end{center}
	\caption{Left panel shows the temperature profile at black string horizon with various values of $\delta$ by fixing $b_0 =0.1, b_1=0.7$. Right panel shows the temperature profile  at black string horizon with various values of $b_1$ by fixing $b_0 =0.1, \delta=1$.}\label{fig:TRb}
\end{figure}

Conveniently, let us write temperatures in terms of dimensionless variables as follows:
\begin{eqnarray}
	\bar{T}_{R(b)}
	&=&\frac{r_V^2}{3\times2^{2/3} M}T_{R(b)} =\frac{\left(b_0+2b_1x- 2^{-2/3}x^2\right)(1+x^2\delta)}{4\pi x},\\
	\bar{T}_{R(c)}
	&=&\frac{r_V^2}{3\times2^{2/3} M}T_{R(c)}=\frac{-\left(b_0+2b_1y- 2^{-2/3}y^2\right)(1+y^2\delta)}{4\pi y},\\
    &&r_b=xr_V,\quad
    r_c=yr_V,\quad
    \lambda=\delta\frac{2} {\pi r_V^2}.
\end{eqnarray}
Note that both temperatures are positive for the range $0<r_b<r_c$ or $0<x<2^{2/3}\Big(b_1+\sqrt{b_1^2+\frac{b_0}{2^{2/3}}}\,\Big)$ and $2^{2/3}\Big(b_1+\sqrt{b_1^2+\frac{b_0}{2^{2/3}}}\,\Big)<y<\frac{1}{2^{1/3}}\left(3b_1+\sqrt{6\times2^{1/3}b_0+9b_1^2}\right)$. For the extremal black string, the parameters $x$ and $y$ are equal to  $x=y=2^{2/3}\Big(b_1+\sqrt{b_1^2+\frac{b_0}{2^{2/3}}}\,\Big)$. In order to find the extremum of $\bar{T}_{R(b)}$, one can solve a condition,
\begin{align}
	\frac{d\bar{T}_{R(b)}}{dx}
	=8b_1x\delta+b_0\left(-\frac{2}{x^2}+2\delta\right)-2^{1/3}(1+3x^2\delta)=0\label{eq:dTb}.
\end{align}
For the case of dS solution in which $m_g^2>0$ and $b_1=b_0=0$, the slope of $T_{R(b)}$ is always negative. It implies that system is locally unstable. Note that the slope of the temperature directly infers the sign of the heat capacity which is always negative in this case. For the dRGT with R\'{e}nyi entropy case, it is possible to find the extrema of the temperature. This means that it is possible to have the positive heat capacity. From Eq. \eqref{eq:dTb}, there are two positive real roots for the viable range of parameters. The results are lengthy and we do not write explicitly here. However, in order to further analyze the thermodynamic properties, the solution can be approximated by using an assumption as $b_0\ll1$. As a result, the solution can be written as 
\begin{align}
	x_{ex\pm}=\frac{2^{5/3}}{3}b_1\left(1\pm \Delta \right)\pm\frac{ b_0}{b_1\Delta}   \left[1+ \frac{3}{2} \frac{\delta}{\delta_c} \left(1\pm \Delta \right)\right], \quad 
	\Delta = \sqrt{1 -\frac{\delta_c}{\delta}},\quad 
	\delta_c = \frac{3}{8\times2^{1/3} b_1^2}.\label{xpm}
\end{align}
To obtain a real positive value of $x_{ex}$, we need $\delta>\delta_c$. The dRGT black string will be locally stable if the nonextensive effect is large enough, i.e., $\delta>\delta_c=3/(8\times2^{1/3} b_1^2)$. This can be shown explicitly in the left panel in Fig. \ref{fig:TRb}. Note that, in the limit $\lambda\rightarrow 0$ called the GB limit, there are no extrema for temperature. 
As seen in Eq. \eqref{xpm}, the bound in $\delta$ depends on the parameter $b_1$. Therefore, it is possible to express the bound in $b_1$ as $b_{1c}=\sqrt{3/(8\times2^{1/3}\delta_c)}$. The temperature profile for varying $b_1$ is illustrated in the right panel in Fig. \ref{fig:TRb}. Obviously, the bound in $b_1$ is indeed the lower bound implied from the parameter $b_{1c}$ being inversely proportional to the nonextensive parameter $\delta_c$.

\begin{figure}[h!]
	\begin{center}
		\includegraphics[scale=0.5]{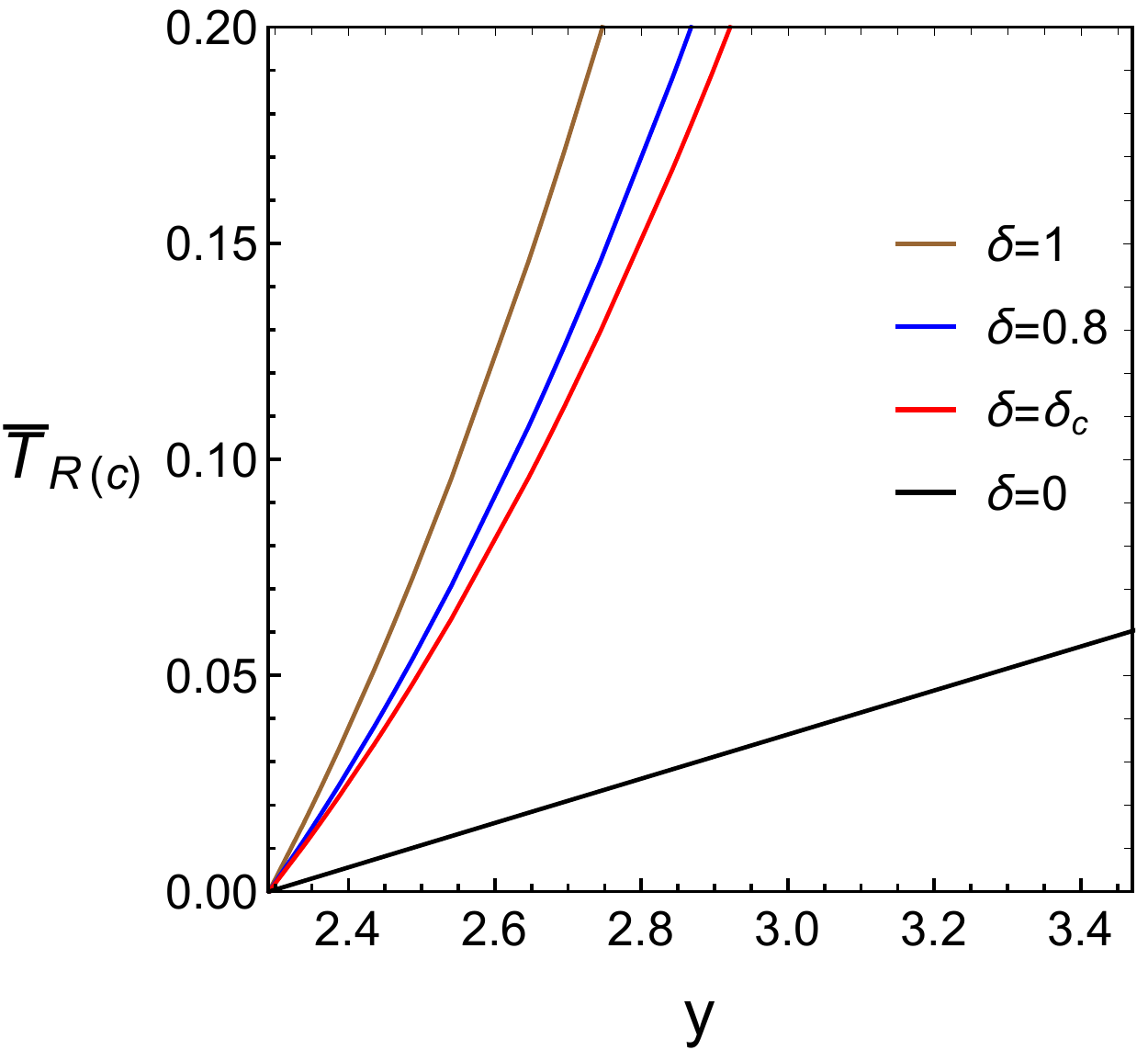}\quad
		\includegraphics[scale=0.5]{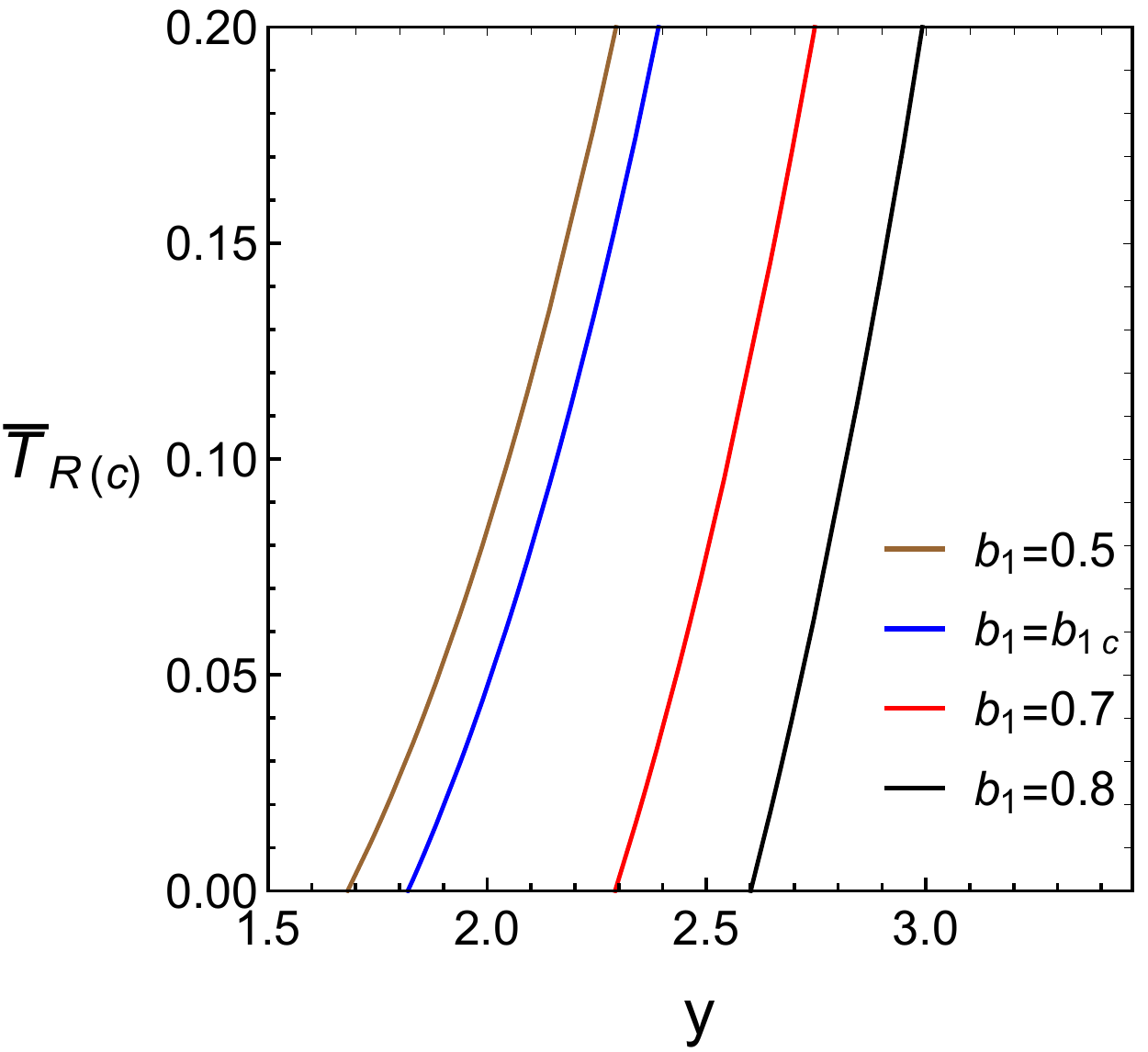}
	\end{center}
	\caption{Left panel shows the temperature profile at cosmic horizon with various values of $\delta$ by fixing $b_0 =0.1, b_1=0.7$ . Right panel shows the temperature profile at cosmic horizon with various values of $b_1$ by fixing $b_0 =0.1, \delta=1$ .}\label{fig:TRc}
\end{figure}

For the thermodynamic system evaluated at the cosmic horizon, the slope of temperature is always positive for $2^{2/3}\Big(b_1+\sqrt{b_1^2+\frac{b_0}{2^{2/3}}}\,\Big)<y<\frac{1}{2^{1/3}}\left(3b_1+\sqrt{6\times2^{1/3}b_0+9b_1^2}\right)$. It is possible to find the extrema  of $\bar{T}_{R(c)}$ by solving $\frac{d\bar{T}_{R(c)}}{dy}=0$. However, these points are out of the range $y$ in our consideration. This implies that there are no extrema for temperature evaluated at cosmic horizon. Moreover, one finds that $\bar{T}_{R(c)}$ and $\frac{d\bar{T}_{R(c)}}{dy}$ are always positive. Therefore, the thermodynamic system evaluated at the cosmic horizon is always locally stable. These behaviors can be shown in the left panel in Fig. \ref{fig:TRc}. In the right panel in this figure, the behaviors for varying $b_1$.

\begin{figure}[h!]
	\begin{center}
		\includegraphics[scale=0.4]{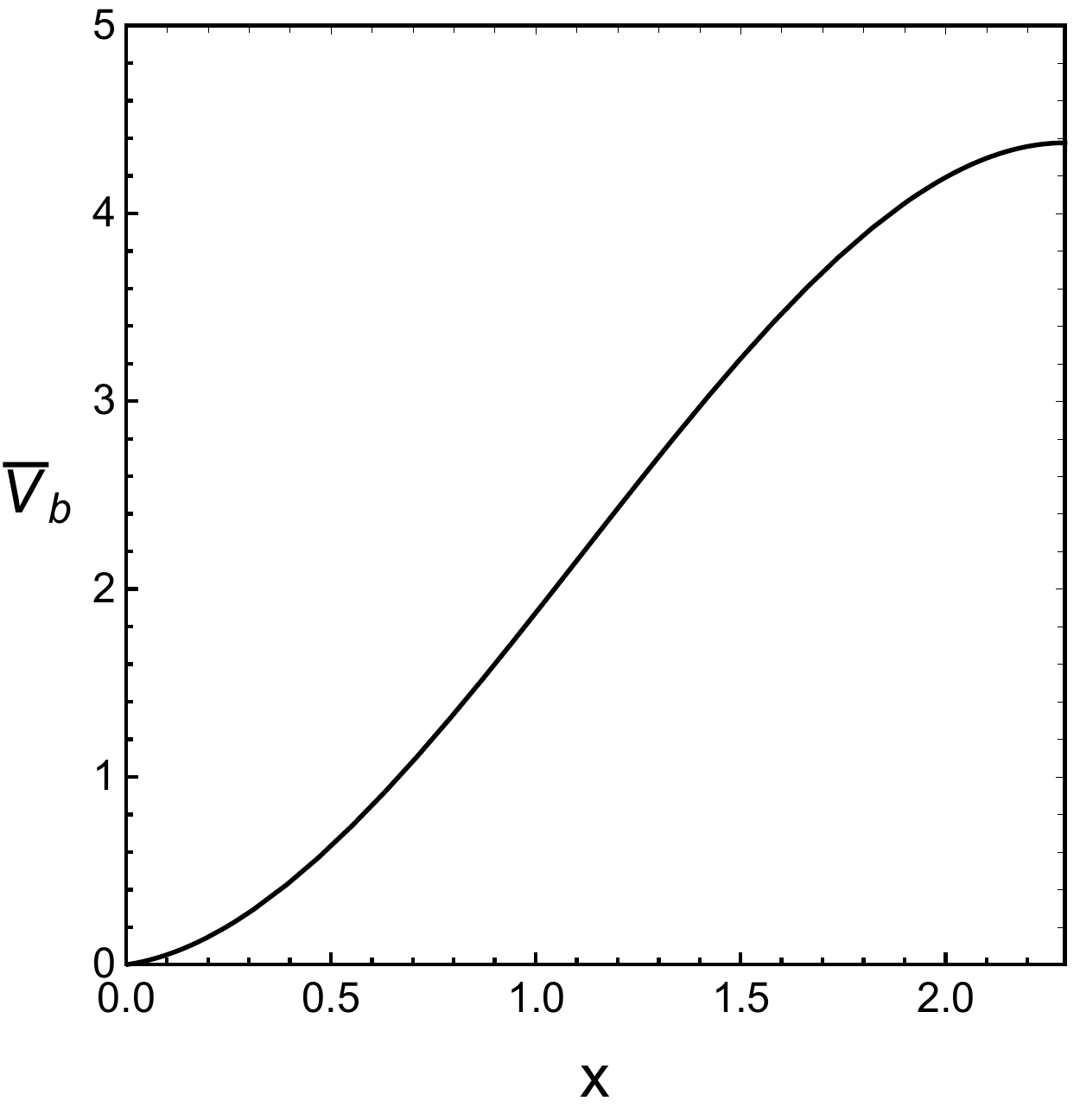}\quad
		\includegraphics[scale=0.4]{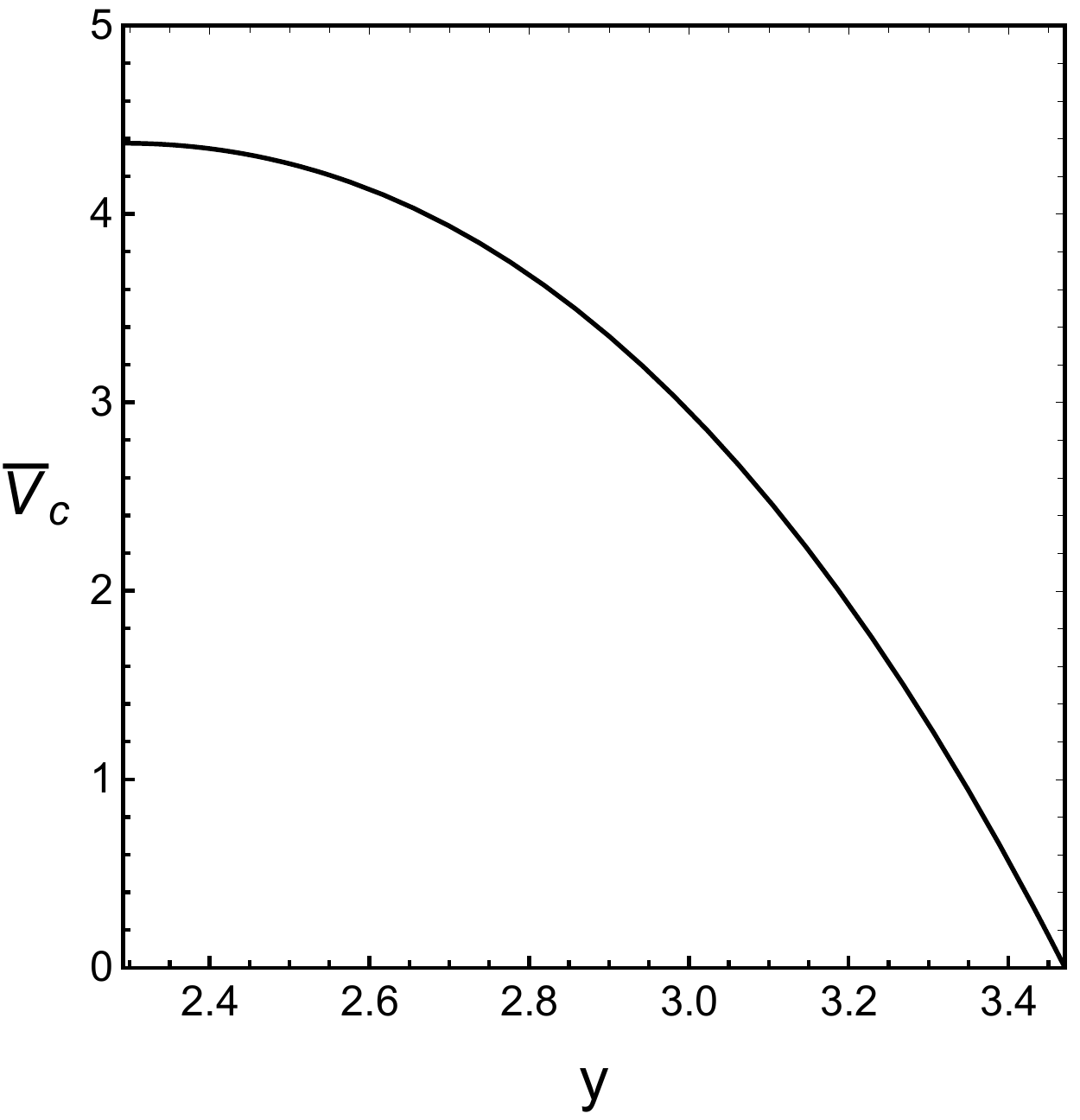}
	\end{center}
	\caption{The volume profile evaluated at black string horizon (left) and cosmic horizon (right) with fixing $b_0 =0.1, b_1=0.7$.}\label{fig:Vbc}
\end{figure}

Let us discuss on the thermodynamic volumes $V_{b,c}$. Their dimensionless expressions read
\begin{eqnarray}
	\bar{V}_b&=&\frac{1}{\pi r_V^3}V_b
	=\frac{2}{3}x^3\left(-1+3\times2^{2/3}\frac{b_1}{x}+3\times2^{2/3}\frac{b_0}{x^2}\right),\\
	\bar{V}_c&=&\frac{1}{\pi r_V^3}V_c
	=\frac{2}{3}y^3\left(-1+3\times2^{2/3}\frac{b_1}{y}+3\times2^{2/3}\frac{b_0}{y^2}\right).
\end{eqnarray}
One can see that these volumes can be negative if the horizon radii are large enough. It is possible to tune the values of $b_0$ and $b_1$ in which both $V_b$ and $V_c$ are positive within their viable ranges as illustrated in Fig. \ref{fig:Vbc}. 

\begin{figure}[h!]
	\begin{center}
		\includegraphics[scale=0.605]{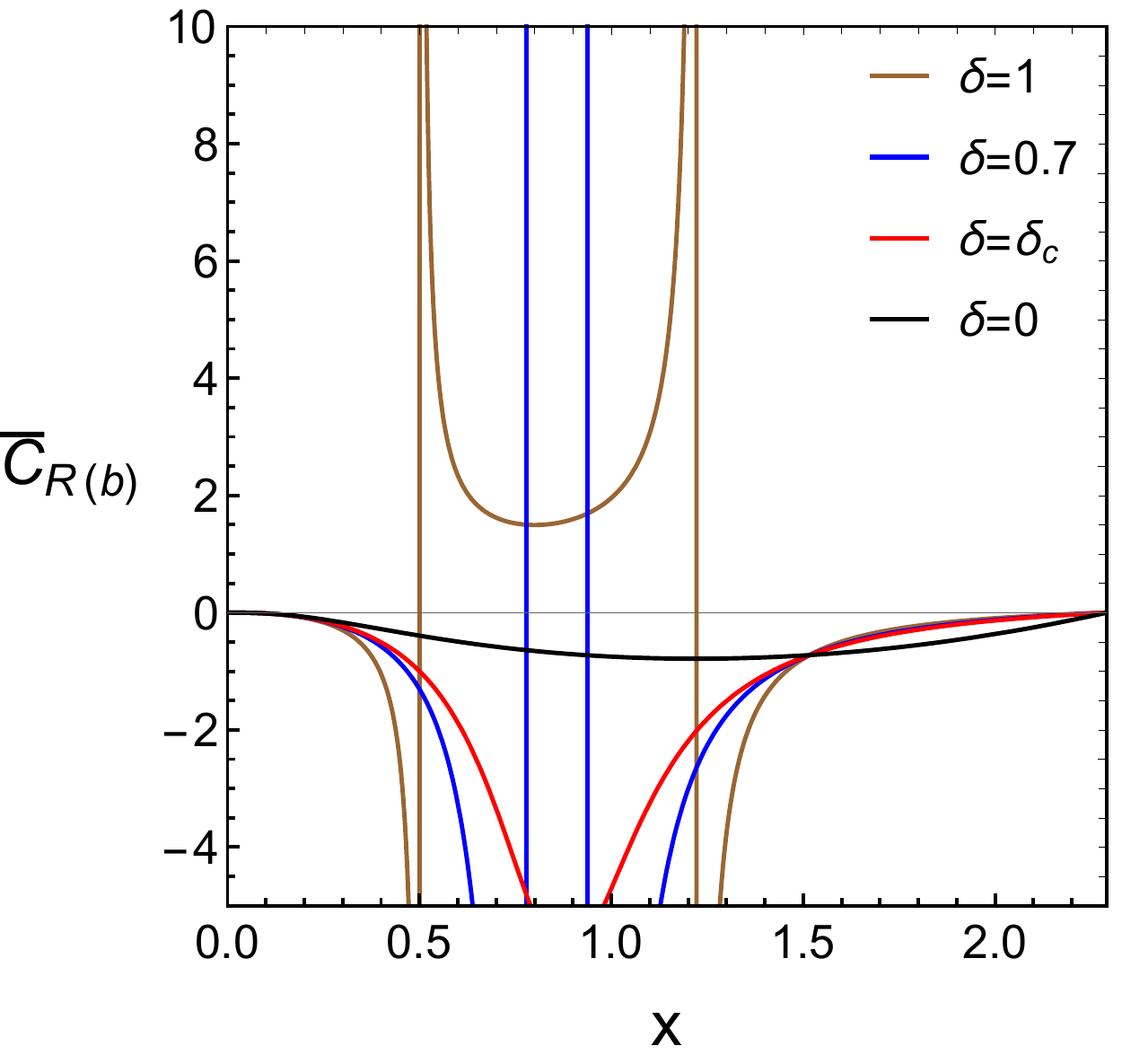}\quad
		\includegraphics[scale=0.61]{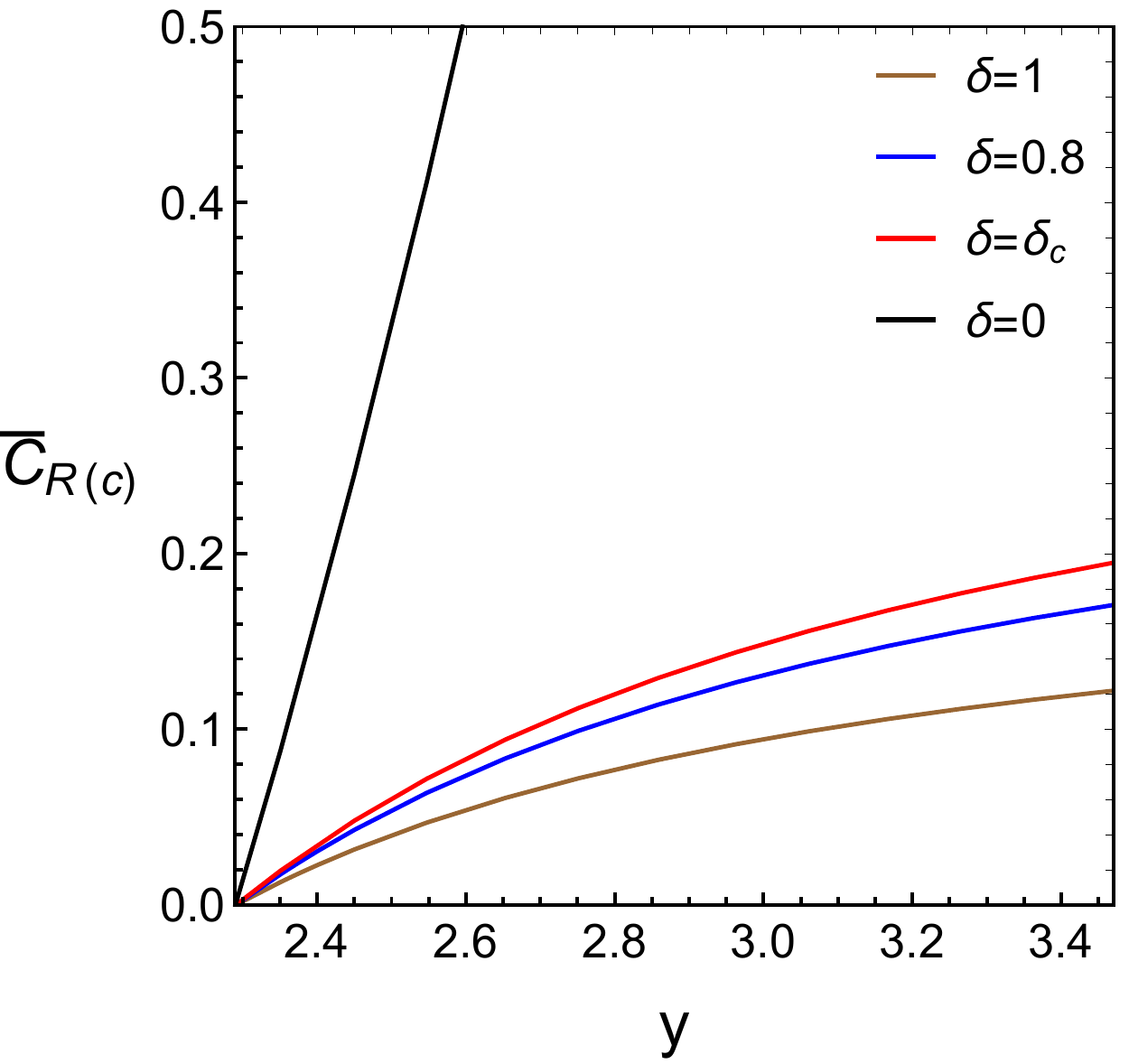}
	\end{center}
	\caption{Left panel shows the heat capacity profile at black string horizon with various values of $\delta$ by fixing $b_0=0.1$, $b_1=0.7$. Right panel shows the heat capacity profile at black string horizon with various values of $\delta$ by fixing $b_0=0.1$, $b_1=0.7$.}\label{fig:Cbc}
\end{figure}

The existence of local minimum temperature is provided locally stable-unstable phase transitions. This behavior can be obtained by analyzing the slope of the temperature, which is proportional to the heat capacity.  For the locally thermodynamic stability, the system is required to have positive heat capacity. As $M$ playing the role of enthalpy, the heat capacity for isobaric process can be written as 
\begin{align}
C_{R(b,c)}&=\pm\Big(\frac{\partial M}{\partial T_{R(b,c)}}\Big)_P,\\
\bar{C}_{R(b)}&=\frac{  x^2(b_0+2b_1x-2^{-2/3}x^2)}{2^{2/3} b_0 \left(\delta  x^2-1\right)-x^2 \left(-4\times 2^{2/3} b_1 \delta  x+3 \delta  x^2+1\right)},\\
\bar{C}_{R(c)}&=\frac{  y^2(b_0+2b_1y-2^{-2/3}y^2)}{2^{2/3} b_0 \left(\delta  y^2-1\right)-y^2 \left(-4\times 2^{2/3} b_1 \delta  y+3 \delta  y^2+1\right)}.
\end{align} 
The behaviors of heat capacity are shown in Fig. \ref{fig:Cbc}. Note that, in this figure, we have used the dimensionless version as $\bar{C}_{R(b)}=\frac{C_{R(b)}}{\pi 2^{2/3}r_V^2}$. The positiveness of the heat capacity can be determined from the positive slope of the temperature. At the black string horizon, there are three ranges of $x$ in which $\bar{C}_{R(b)}$ is positive for the middle part and the others are negative. Moreover, $\bar{C}_{R(b)}$ changes the sign at extremum points of temperature. As a result, the moderate-sized black string is locally stable while the smaller and larger ones are locally unstable. In other words, there are three possible states of black string. The small black string with higher temperature will radiate thermal energy then it will eventually evaporate away, while that with the lower temperature will evolve to the moderate-sized black string which lies on the ranges of positive $\bar{C}_{R(b)}$. The large black string with higher temperature will evaporate as losing its mass through thermal radiation and then becomes to moderate-sized eventually. Therefore, the moderated-sized black string state is more stable than the other states.

For the heat capacity evaluated at the cosmic horizon, there are no divergent points for $\bar{C}_{R(c)}$, since there are no extrema of $\bar{T}_{R(c)}$ for the range $2^{2/3}\Big(b_1+\sqrt{b_1^2+\frac{b_0}{2^{2/3}}}\,\Big)<y<\frac{1}{2^{1/3}}\left(3b_1+\sqrt{6\times2^{1/3}b_0+9b_1^2}\right)$. Therefore, there is no a phase transition for the system evaluated at $r_c$. Moreover, one finds that the heat capacity of this system is always positive. This is also compatible to the slope of the temperature that we have analyzed. Note that, in the GB limit, it is not possible to have positive value of the heat capacity as shown in the black line in the right panel in Fig. \ref{fig:Cbc}.

\begin{figure}[h!]
	\begin{center}
		\includegraphics[scale=0.6]{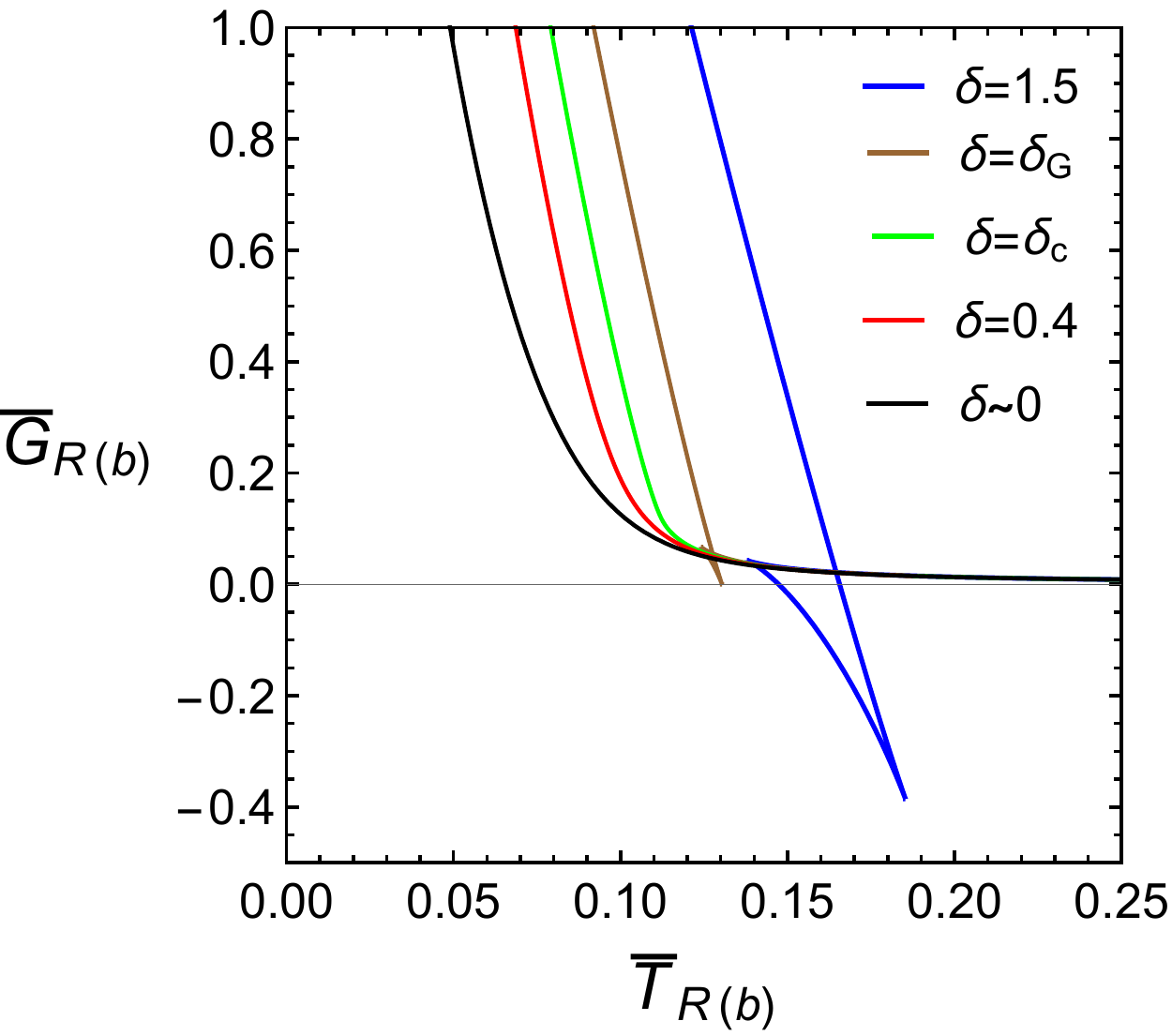}\quad
		\includegraphics[scale=0.65]{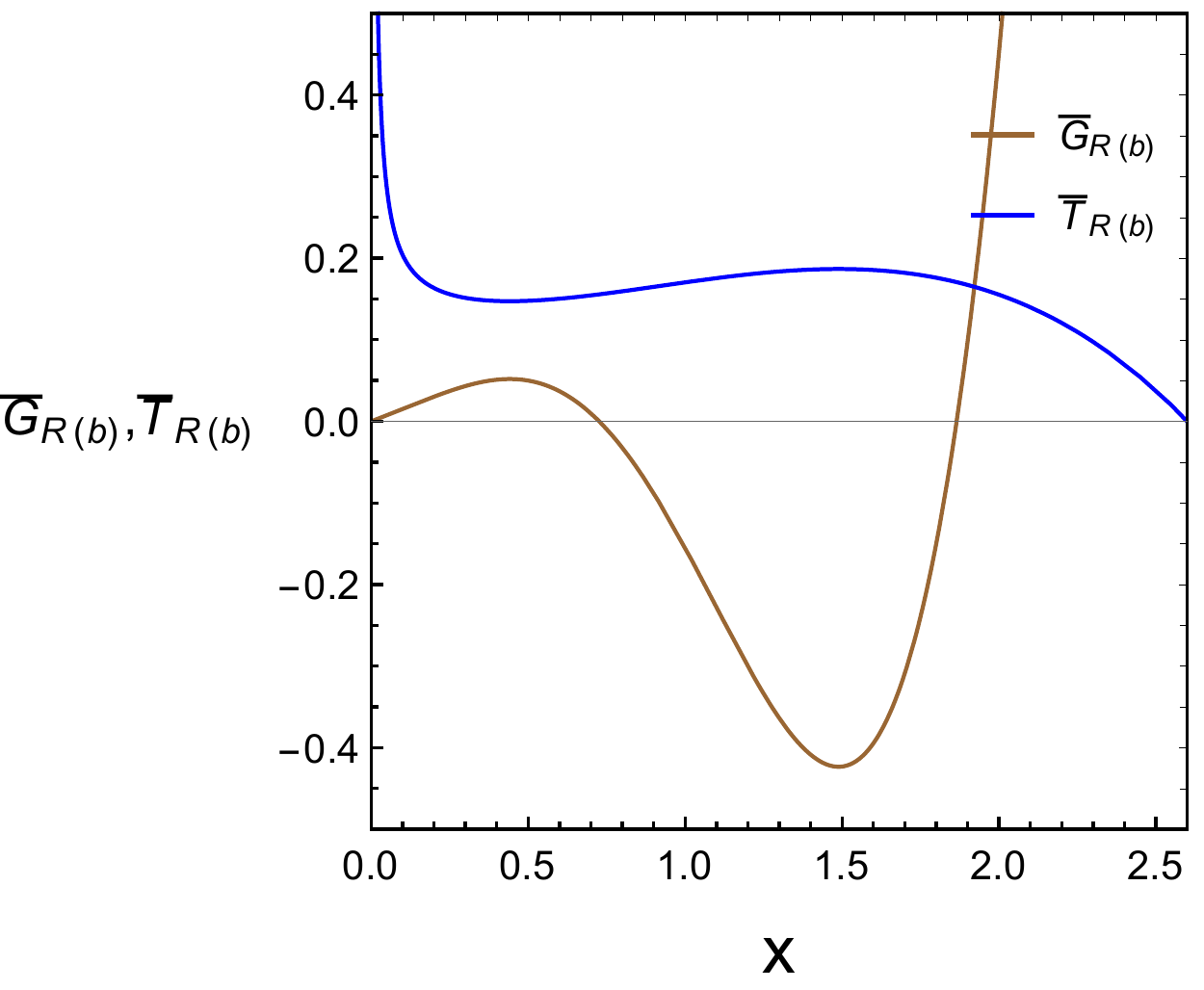}
	\end{center}
	\caption{Left panel shows the Gibbs free energy against the temperature at black string horizon with various values of $\delta$ by fixing $b_0=0.1$, $b_1=0.7$. Right panel shows the Gibbs free energy and the temperature with respect to $x$ by fixing $b_0=0.1$, $b_1=0.8$ and $\delta=1$.}\label{fig:GP}
\end{figure}

We have already analyzed the local stability of the black string by considering the behavior of the heat capacity. Now, we will investigate the global stability by considering the Gibbs free energy. The system with lower Gibbs free energy at a given temperature prefers to exist compared to those with higher free energy. This state is called being globally stable. For example, if the free energy of the system without black string is zero, thus black string can be formed by the condition $G<0$. The Gibbs free energy can be expressed as 
\begin{align}
G_{R(b,c)}&=M-T_{R(b,c)}S_{R(b,c)},\\
\bar{G}_{R(b)}=\frac{G_{P(b)}}{2^{-4/3}M}&=x\left(3b_0+3b_1-\frac{x^2}{2^{2/3}}\right)-\frac{3(b_0+2b_1x-2^{-2/3}x^2)(1+x^2\delta)\ln(1+x^2\delta)}{2x\delta},\label{GPb}\\
\bar{G}_{R(c)}=\frac{G_{P(c)}}{2^{-4/3}M}&=y\left(3b_0+3b_1-\frac{y^2}{2^{2/3}}\right)+\frac{3(b_0+2b_1y-2^{-2/3}y^2)(1+y^2\delta)\ln(1+y^2\delta)}{2y\delta}.
\end{align}
Notice that the entropy of black string is always positive so that slope of the graph $G_R-T_R$ is always negative, since $\Big(\frac{\partial G_{R(b,c)}}{\partial T_{R(b,c)}}\Big)_P=-S_{R(b,c)}$. This implies that if entropy keeps increasing, the slope will be more negative. The behavior of Gibbs free energy against the temperature is shown in the left panel in Fig. \ref{fig:GP}. There exist two cusps corresponding to two extrema in the temperature profile denoted by $x_{ex\pm}$. Note that the slopes of the Gibbs free energy at these points are still continuous. These points also correspond to the locally stable-unstable phase transitions, since $\Big(\frac{\partial^2 G_{R(b)}}{\partial T_{R(b)}^2}\Big)_P\propto C_{R(b)}$, where the heat capacity diverges and changes the sign at these points. This is also shown explicitly in the right panel in Fig. \ref{fig:GP}. The local maximum/minimum of the Gibbs free energy is at the same point with the minimum/maximum of the temperature. We can see that, for the range from $x=0$ to $x=x_{ex-}$, the Gibbs free energy increases as temperature decreases, while the Gibbs free energy decreases as the temperature increases for the range from $x=x_{ex-}$ to $x=x_{ex+}$, and lastly the Gibbs free energy will increases with temperature decreasing again for the range from $x=x_{ex+}$ to $x=2^{2/3}\Big(b_1+\sqrt{b_1^2+\frac{b_0}{2^{2/3}}}\,\Big)$. According to this result, one can see that it is possible to obtain the globally stable black string with the dimensionless horizon radius between $x=x_{ex-}$ to $x=x_{ex+}$, since there is a part with negative free energy at a given temperature. 

\begin{figure}[h!]
	\begin{center}
		\includegraphics[scale=0.55]{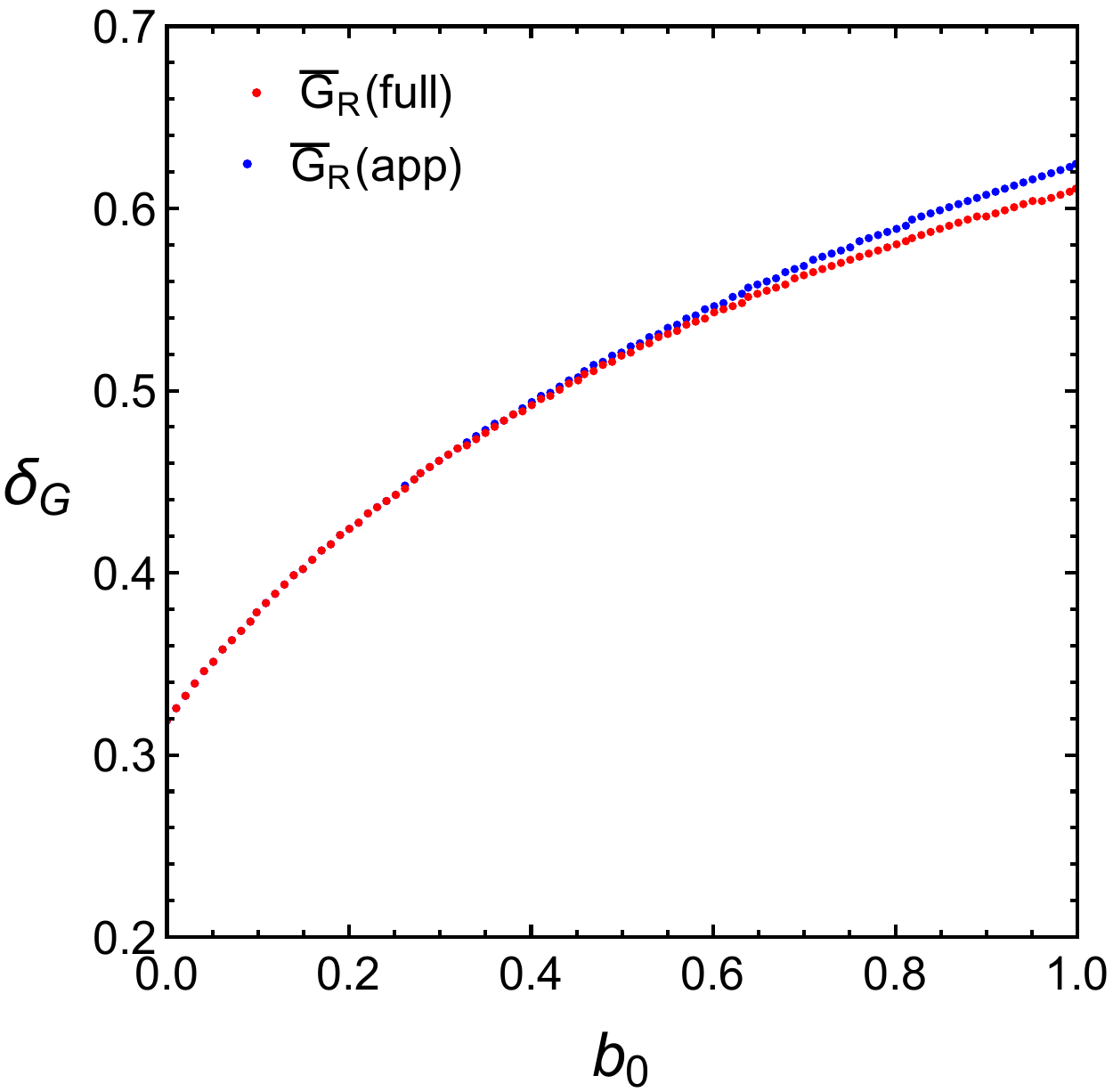}\quad
		\includegraphics[scale=0.55]{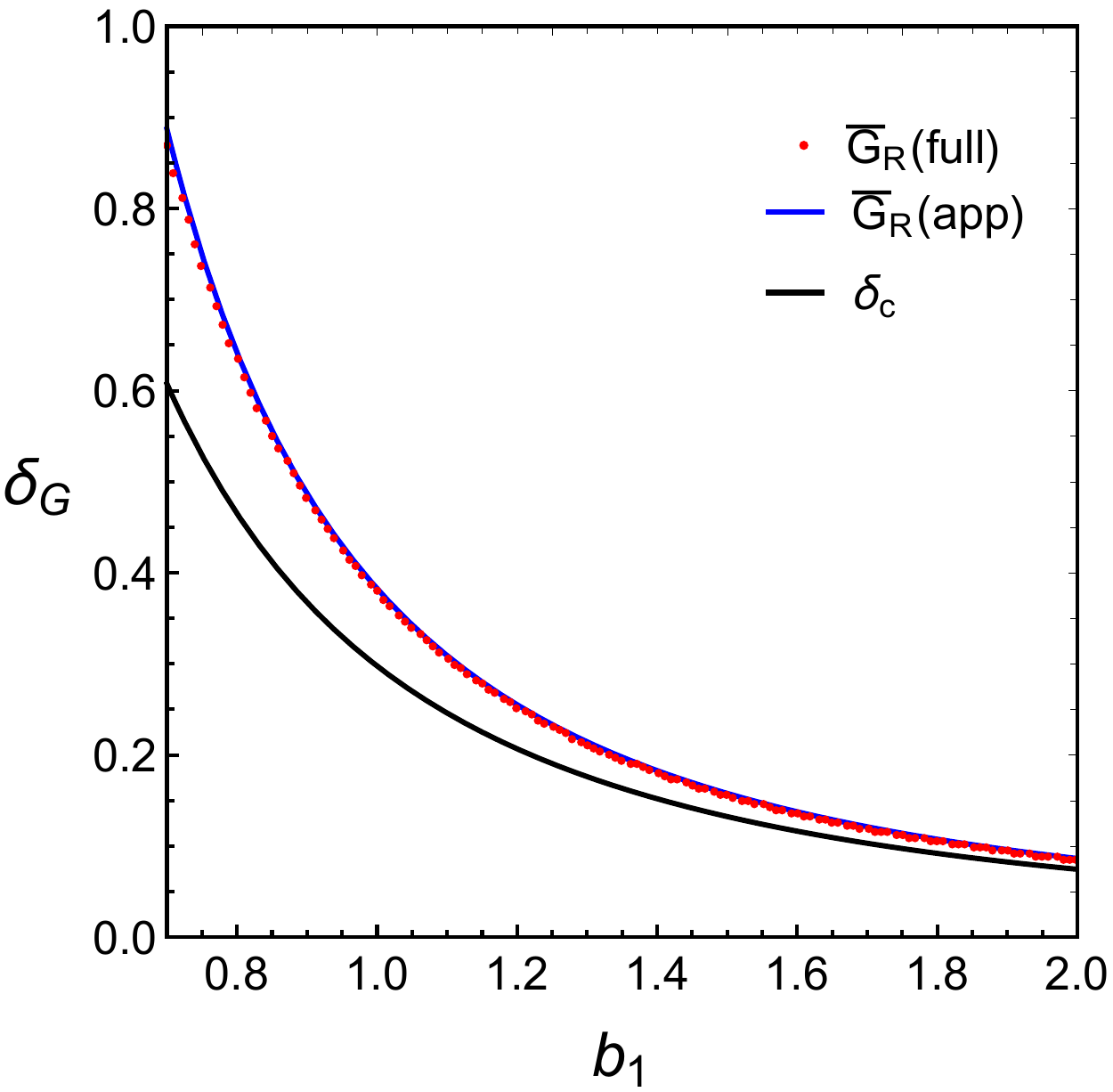}
	\end{center}
	\caption{Left panel shows the comparison of $\delta_G$ obtained from $\bar{G}_R(b)$ in full equation and approximation by fixing $b_1=1$. Right panel shows the comparison of $\delta_G$ obtained in full expression and approximation by fixing $b_0=1$. The black solid curve represents the bound for the local stability $\delta_c$}\label{fig:delta-ep}
\end{figure}

Moreover, from the left panel in Fig. \ref{fig:GP}, we can see that there exists a critical value of $\delta$ such that it is not possible to obtain negative Gibbs free energy. Therefore, the lower bound for $\delta$ can be found by requiring the condition $\bar{G}_{R(b)}|_{x_{ex+}}<0$. This value will be denoted by $\delta_G$. In principle, one can find the expression of $\delta_G$ in terms of parameters $b_0$ and $b_1$, since $x_{ex+}$ depends on $b_0, b_1$ and $\delta$. However, the expression is very lengthy and it is not suitable to show explicitly. In order to obtain the analytical expression of $\delta_G$, we use numerical method evaluating point by point to show that the bound $\delta_G$ slightly depends on the parameter $b_0$ as shown in the left  panel in Fig. \ref{fig:delta-ep}. From this figure, one can see that  the approximated value of the bound is still trustable for $b_0 \ll 1$. Therefore, one can use the approximation $b_0 \sim 0$ in order to find the analytic expression of $\delta_G$. By substituting $x_{ex+}$ from Eq. \eqref{xpm} to $\bar{G}_{R(b)}$ in Eq. \eqref{GPb} and then using approximation $b_0 \ll 1$ and $\Delta \sim 0$ (the approximated free energy denotes as $\bar{G}_{R(\text{app})}$), the bound can be expressed as 
\begin{align}
    \delta_G
    =\frac{9}{8}\frac{\left[e^{\left\{{\frac{7}{6}+PL\left(-\frac{7}{6e^{7/6}}\right)}\right\}}-1\right]}{2^{1/3}b_1^2}+\frac{b_0}{2b_1^4}
    =3\left[e^{\left\{{\frac{7}{6}+PL\left(-\frac{7}{6e^{7/6}}\right)}\right\}}-1\right]\delta_c+\frac{b_0}{2b_1^4},
\end{align}
where $PL(z)$ is the ProductLog function. This function returns the value of $x$ by solving $z=xe^x$. 
From the right panel in Fig. \ref{fig:delta-ep}, one can see that the bound from the above expression is very closed and slightly greater than to the numerical result. From this figure, one also see that it is in the same shape with $\delta_c$ but stronger than $\delta_c$.

In order to have both locally and globally thermodynamic stability, the system at black string horizon must be satisfied the condition $\delta>\delta_G$. It is important to note that there exits a point such that the Gibbs free energy of the non-black string state (or hot gas state) and black string state are the same. At this point, the hot gas will transform to the moderate-sized black string. Since the slope of the Gibbs free energy at the transition point is discontinuous, the transition is the first-order phase transition commonly known as the Hawking-Page phase transition. For the system evaluated at cosmic horizon, there are no extrema for $T_{R(c)}$. Therefore, there are no cusps as found in one evaluated at black string horizon. Moreover, $\bar{G}_{R(c)}$ is always negative, and then the black string is stable for the range of $\delta>\delta_G$.


\section{effective thermodynamic systems}\label{sect: Effective thermo}

\begin{figure}[h!]
	\begin{center}
		\includegraphics[scale=0.55]{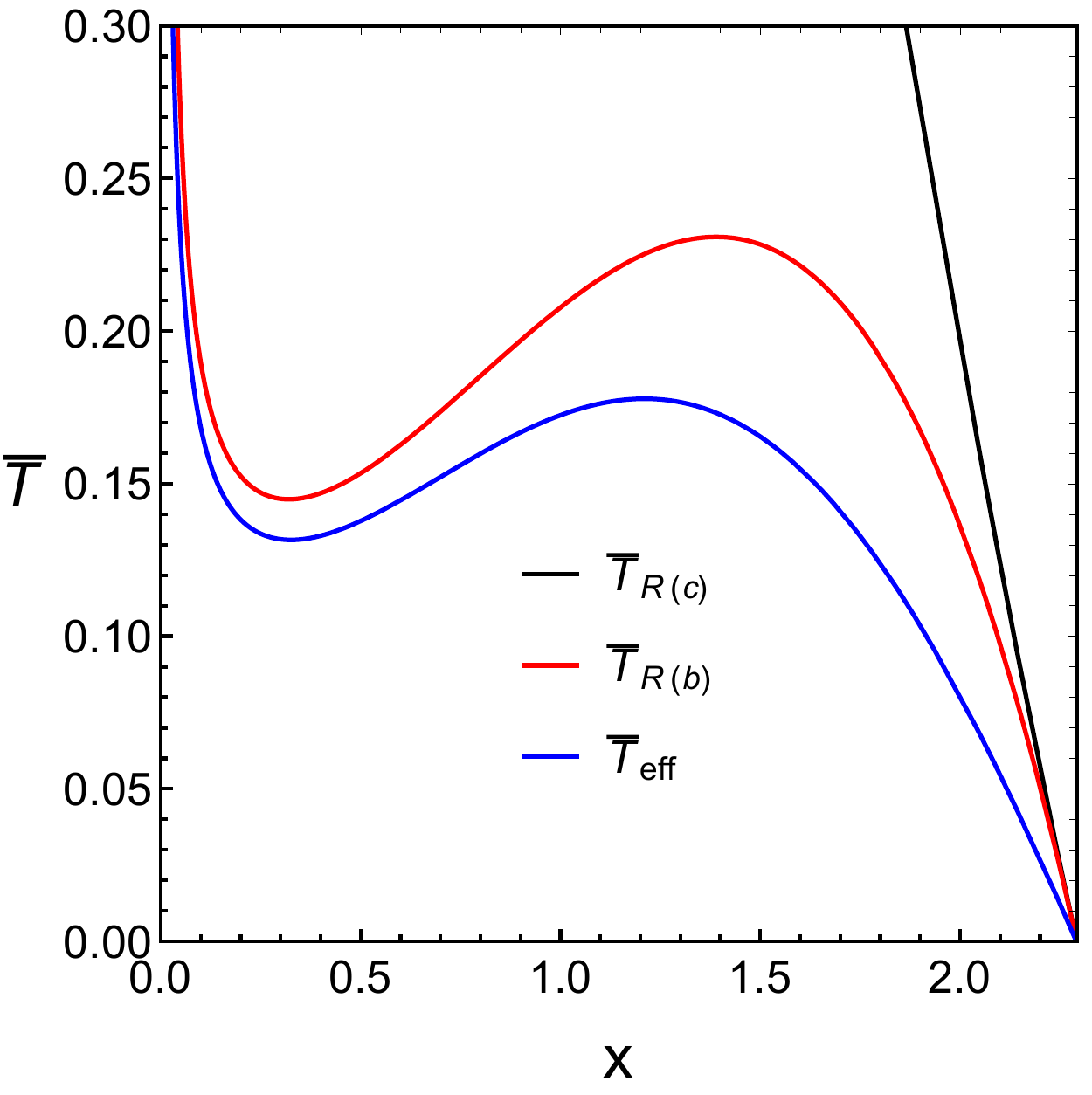}\quad
		\includegraphics[scale=0.364]{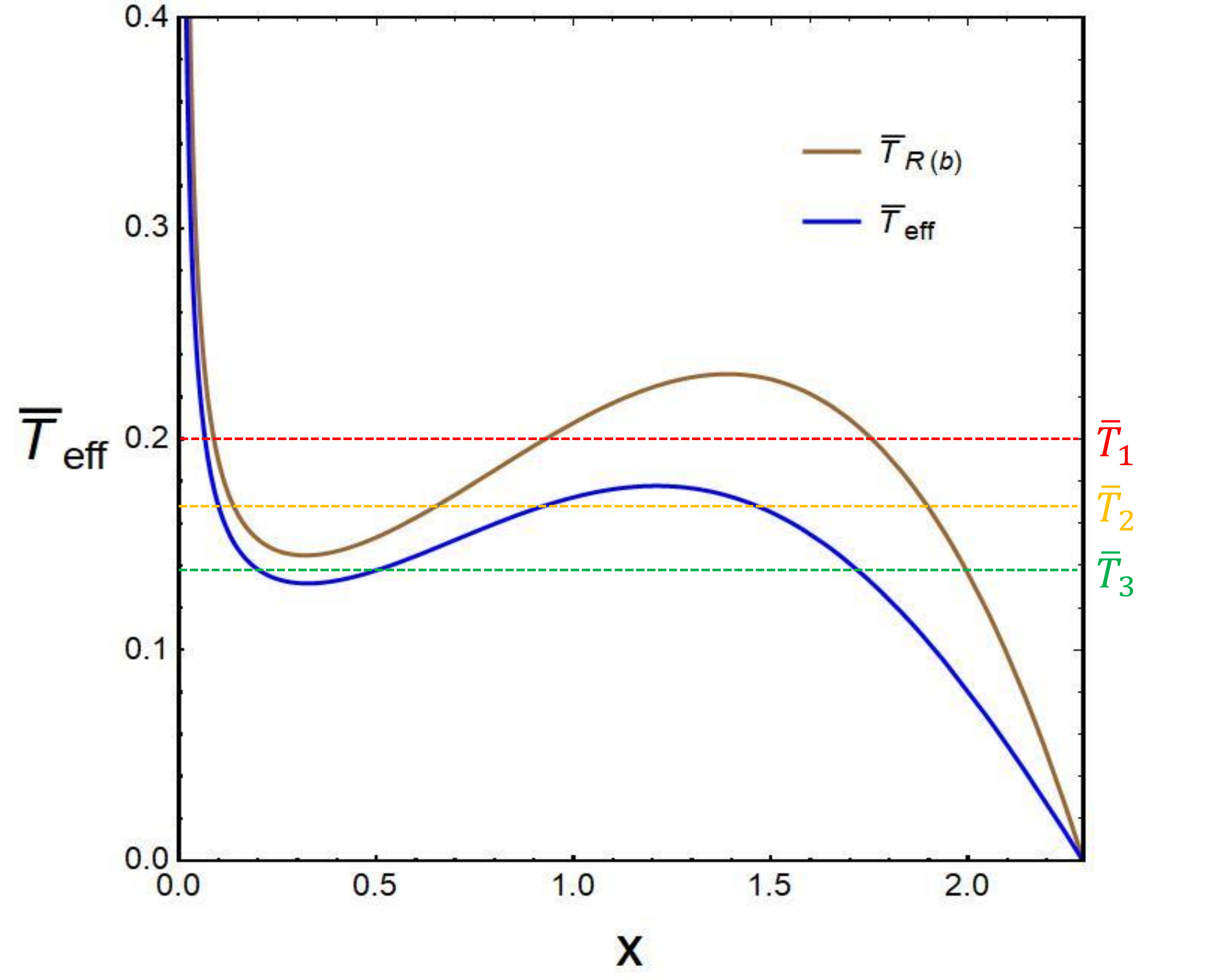}
	\end{center}
	\caption{ Left panel shows the comparison of temperatures of the system evaluated at black string horizon, cosmic horizon and effective system by fixing $b_0 =0.1, b_1=0.7$ and $\delta=2$. Right panel shows the comparison of temperatures of the system evaluated at black string horizon and effective system by fixing $b_0 =0.1, b_1=0.7$ and $\delta=2$}\label{fig:Teff}
\end{figure}

For the effective system approach, the whole system is regarded as a single system. The entropy of the effective system is supposed to be an addition of those of two systems as
\begin{align}
	S=S_{R(b)}+S_{R(c)}
	&=\frac{1}{\lambda}\ln\left(1+\lambda \frac{\pi r_b^2}{2}\right)+\frac{1}{\lambda}\ln\left(1+\lambda \frac{\pi r_c^2}{2}\right)\nonumber\\
	&=\frac{1}{\lambda}\ln\left[\left(1+\lambda \frac{\pi r_b^2}{2}\right)\left(1+\lambda \frac{\pi r_c^2}{2}\right)\right].
\end{align}
To obtain the real value of the effective entropy, it requires the condition as
\begin{align}
	\left(1+\lambda \frac{\pi r_b^2}{2}\right)\left(1+\lambda \frac{\pi r_c^2}{2}\right)>0.
\end{align} 
To satisfy this condition, it is sufficient to restrict our consideration on the positive value of $\lambda$, $\lambda > 0$. For the effective system, it is possible to treat the variable $M$ as the internal energy and the enthalpy. In this work, we choose to consider the parameter $M=M(S,P)$ as the enthalpy of the system, since this choice allows us to compare the result with the separated system approach. As a result, the first law for the effective system approach can be written as
\begin{align}
	dM=T_\text{eff}dS+V_\text{eff}dP,
\end{align}
where the pressure of this effective system is also defined as the same as one in separated system approach, i.e., $P=\frac{3}{8\pi}m_g^2$. For the effective quantities such as $T_\text{eff}$ and $V_\text{eff}$, we follow the notion from Ref. \cite{Nakarachinda:2021jxd}. From this notion, the sign of heat transfer for the system evaluated at the cosmic horizon is opposite to one at black string horizon. This is due to the fact that the observer stays between both horizons. Moreover, the resulting quantities match with the first law of thermodynamics. As a result, the effective temperature is expressed as
\begin{align}
	T_\text{eff}
	&=\Big(\frac{\partial M}{\partial S}\Big)_P
\nonumber,\\
	&=\frac{m_g^2\Big[r_c \left(2 c_1-3 r_c\right)+c_0\Big]\Big[r_b\left(2 c_1-3 r_b\right)+c_0\Big] \left(\pi  \lambda  r_b^2+2\right) \left(\pi  \lambda  r_c^2+2\right)}{8 \pi  \left(r_b-r_c\right)
	\left(c_0\left(2-\pi\lambda r_br_c\right)
	+r_br_c\Big[3\Big\{\pi\lambda\left(r_b r_c+r_b^2+r_c^2\right)+2\Big\}
	-2\pi c_1\lambda\left(r_b+r_c\right)\Big]\right)}.\label{Eq:Teff}
\end{align}
This form of the effective temperature can be written in terms of $T_{R(b)}$ and $T_{R(c)}$ as follows:
\begin{align}
	\frac{1}{T_\text{eff}}=\left(\frac{\partial S_{R(b)}}{\partial M}\right)_P-\left(\frac{\partial S_{R(c)}}{\partial M}\right)_P=\frac{1}{T_{R(b)}}+\frac{1}{T_{R(c)}}. \label{eq:Teffrbrc}
\end{align}
It is important to note that the effective temperature can be reduced to the black string temperature for the limit $r_c\rightarrow\infty$ and to one for cosmic horizon by taking $r_b\rightarrow0$,
\begin{align}
	\lim_{r_c\to\infty}T_\text{eff}
	&=T_{R(b)}=\frac{m_g^2\left(c_0+2c_1 r_b - 3 r^2_b\right)}{4\pi r_b}\left(1+\lambda \frac{\pi r_b^2}{2}\right),\\
	\lim_{r_b\to 0}T_\text{eff}
	&=T_{R(c)}=-\frac{m_g^2\left(c_0+2c_1 r_c - 3 r^2_c\right)}{4\pi r_c}\left(1+\lambda \frac{\pi r_c^2}{2}\right).
\end{align} 
Since this temperature depends on $r_b$ and $r_c$ which are independent, the slope of the temperature profile does not imply the sign of the heat capacity as we have analyzed in previous section. This is due to the fact that the heat capacity is evaluated as the change of temperature with fixing the pressure. Hence, in order to obtain the temperature profile satisfying the behavior of the heat capacity, we have to find relation between $r_b$ and $r_c$ such that the pressure is held fixed. Therefore, one can find the relation from $r_c\mathcal{F}(r_c)-r_b\mathcal{F}(r_b)=0$. As a result, $r_c$ can be written in terms of $r_b$ as  
\begin{align}
	r_c=\frac{\sqrt{2c_1 r_b-3r_b^2+c_1^2+4 c_0}+ r_b-c_1}{2}.\label{y+}
\end{align}
Substituting  $r_c$ from the above equation to $T_\text{eff}$ in Eq. \eqref{Eq:Teff}, then we obtain $T_\text{eff}=T_\text{eff} (r_b)$. The expression is lengthy so that we do not show here. It is more convenient to use the numerical plot to see the behavior as shown in the Fig. \ref{fig:Teff}. Note that we have used the dimensionless variables in this plot as $\bar{T}_\text{eff}=\frac{r_V^2}{3\times2^{2/3}M}T_\text{eff}$. From the left panel in this figure, one can see that there exists the positive slope of the temperature. This implies that there is the suitable size of the black string corresponding to the positive heat capacity. The locus is similar to one for the system evaluated at the black string horizon. It is also seen that the effective temperature is always less than one evaluated at the black string horizon which is compatible with formula \eqref{eq:Teffrbrc}. As a result, at certain size of the stable black string, the effective temperature is always less than one evaluated at the black string horizon. Moreover, from the right panel in Fig. \ref{fig:Teff}, there exists a particular low temperature (e.g. $\bar{T}_3$) at which only the black string in effective system approach will be locally stable while one for the separated system approach is not. For the same argument, there exists a particular high temperature (e.g. $\bar{T}_1$) at which only the black string in effective system approach will not be locally stable while one for the separated system approach is locally stable.  Moreover, for a particular temperature (e.g. $\bar{T}_2$) at which the systems in both approaches are locally stable, the black string from the effective system approach is always larger than one in the separated system approach. Therefore, this criteria provides us how to distinguish the thermodynamic description for the dRGT black string if this black string really exists in nature.

\begin{figure}[h!]
	\begin{center}
		\includegraphics[scale=0.6]{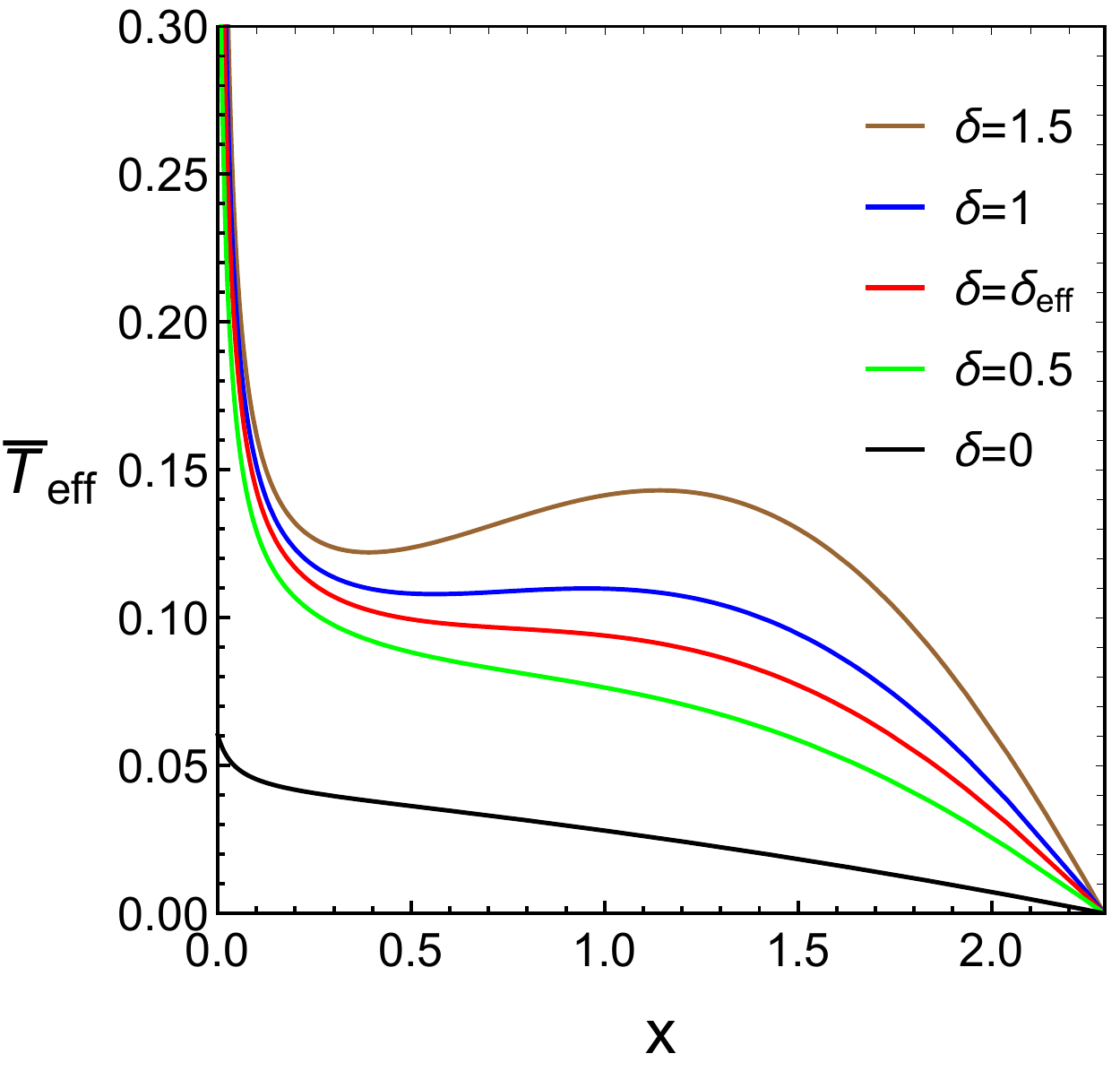}
	\end{center}
	\caption{ The effective temperature profile with various values of $\delta$ by fixing $b_0 =0.1, b_1=0.7$.}\label{fig:TeffR}
\end{figure}

Let us consider the effect of nonextensive parameter on the temperature profile for the effective system approach. Since it has the similar locus as one in the separated system approach, it is possible to exist the lower bound of the nonextensive parameter as shown in Fig. \ref{fig:TeffR}. In order to find the bound, one can use the same strategy as performed in the previous section by finding the condition to have a positive real solution of the equation $\partial_{r_b} T_{R(b)} = 0$. However, for the effective case, the temperature depends on both $r_b$ and $r_c$. Therefore, one has to find the condition for the existence of the extrema along direction with fixing pressure. As a result, the equation for the effective system approach can be written as
\begin{align}
	F(r_b,r_c) =\partial_{r_b} T_{R(b)}+H(r_b,r_c)=0,\quad 
	H(r_b,r_c)=\partial_{r_c}T_{R(c)}\frac{T_{R(b)}^2}{T_{R(c)}^2}\frac{d r_c}{d r_b}. \label{eq:con-d-eff}
\end{align}
As we have analyzed previously, $\partial_{r_b} T_{R(b)}$ is a convex function and depends only on $r_b$. The additional function $H(r_b,r_c)$ is always negative, since $\partial_{r_c}T_{R(c)} > 0$ and $\frac{d r_c}{d r_b}<0$. Moreover, since the function $H(r_b,r_c)$ has a part which is divided by $T_{R(c)}^2$, it will be a small function. Note that $T_{R(c)}$ is much greater than $T_{R(b)}$ as found in the left panel in Fig. \ref{fig:Teff}. Therefore, the function $F(r_b,r_c)$ can be written as a convex function subtracted by a small positive function. By using Eq. \eqref{y+}, the function, $F(r_b,r_c)$, can be written as a function of only $r_b$. Therefore, in principle, one can find the condition on $\delta$ to satisfy the existence of the solution of the above equation, since it is the convex function. By using numerical method, one can find the bound on the nonextensive parameter as 
\begin{align}
	\delta_\text{eff}\geq1.254\left(\frac{3}{8\times2^{1/3} b_1^2}\right)= 1.254 \delta_c.
\end{align}
One can see that the bound from the effective system approach is stronger than one in the separated system approach, $\delta_\text{eff} > \delta_c$. This also infers from Eq. \eqref{eq:con-d-eff}, since the convex function for effective system approach is lower than one in separated system approach. The behavior of the temperature profile with various values of $\delta$ is illustrated in Fig. \ref{fig:TeffR}.

Before discussing on the local stability, let us consider the effective volume which is computed from \cite{Nakarachinda:2021jxd}
\begin{eqnarray}
	V_\text{eff}=T_\text{eff}\left(\frac{V_b}{T_{R(b)}}+\frac{V_c}{T_{R(c)}}\right).
\end{eqnarray}
It can be realized that $V_c=V_c(r_c)$ is indeed the same function as $V_b=V_b(r_b)$, just replacing the different range of the horizon radius, $r_c\to r_b$. Hence, using this fact and Eq. \eqref{eq:Teffrbrc}, the effective is identical to $V_b$ for the variable $r_b$ and to $V_c$ for the variable $r_c$,
\begin{eqnarray}
	V_\text{eff}(r_b)=V_b(r_b),\quad 
	V_\text{eff}(r_c)=V_c(r_c).
\end{eqnarray}
It is then possible to choose suitable values of the parameters $b_0$ and $b_1$ corresponding to the effective volume being positive within its viable range. One also notes that this effective volume is independent of the nonextensive parameter $\lambda$ or $\delta$.

\begin{figure}[h!]
	\begin{center}
		\includegraphics[scale=0.55]{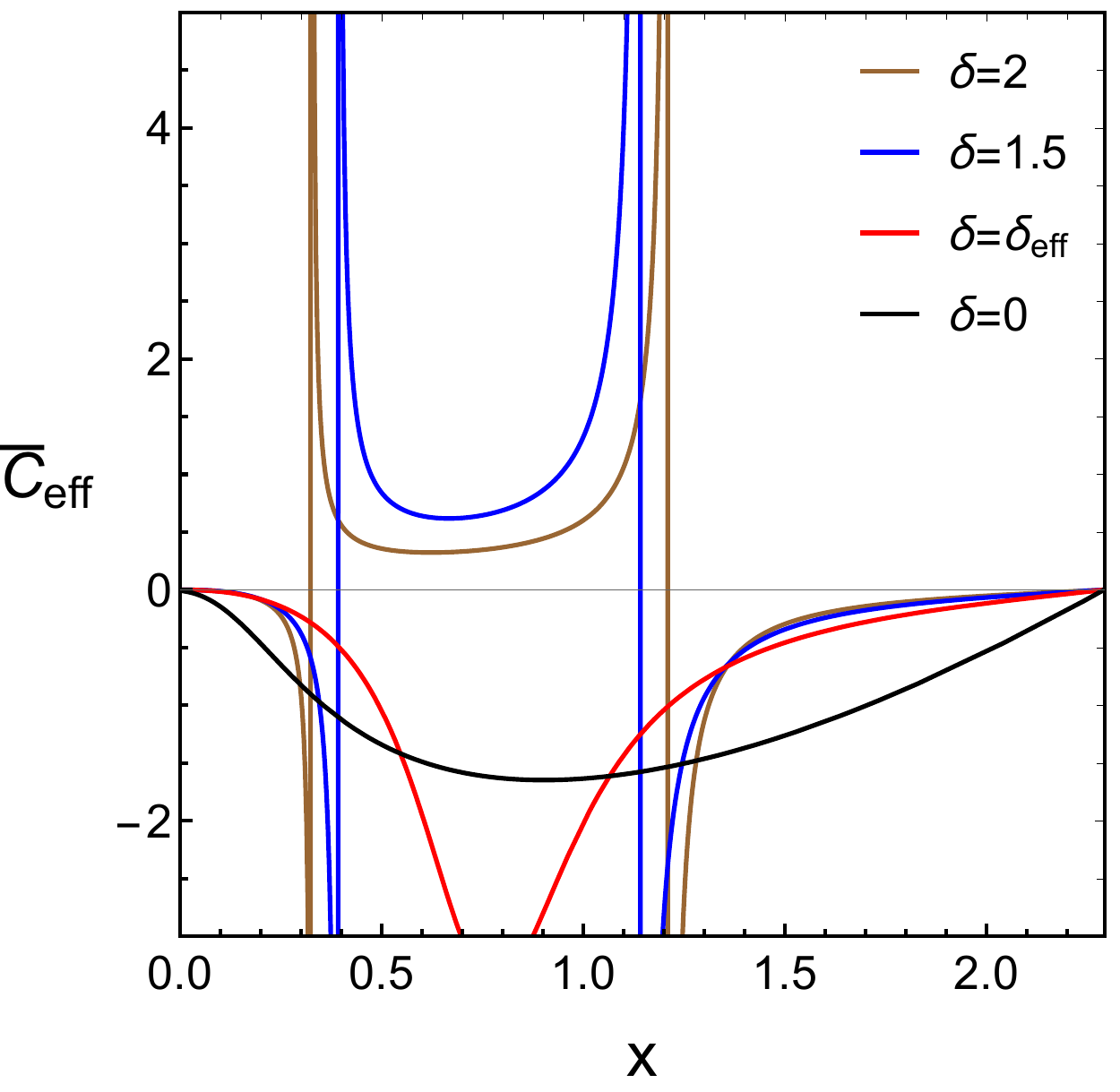}\quad
		\includegraphics[scale=0.58]{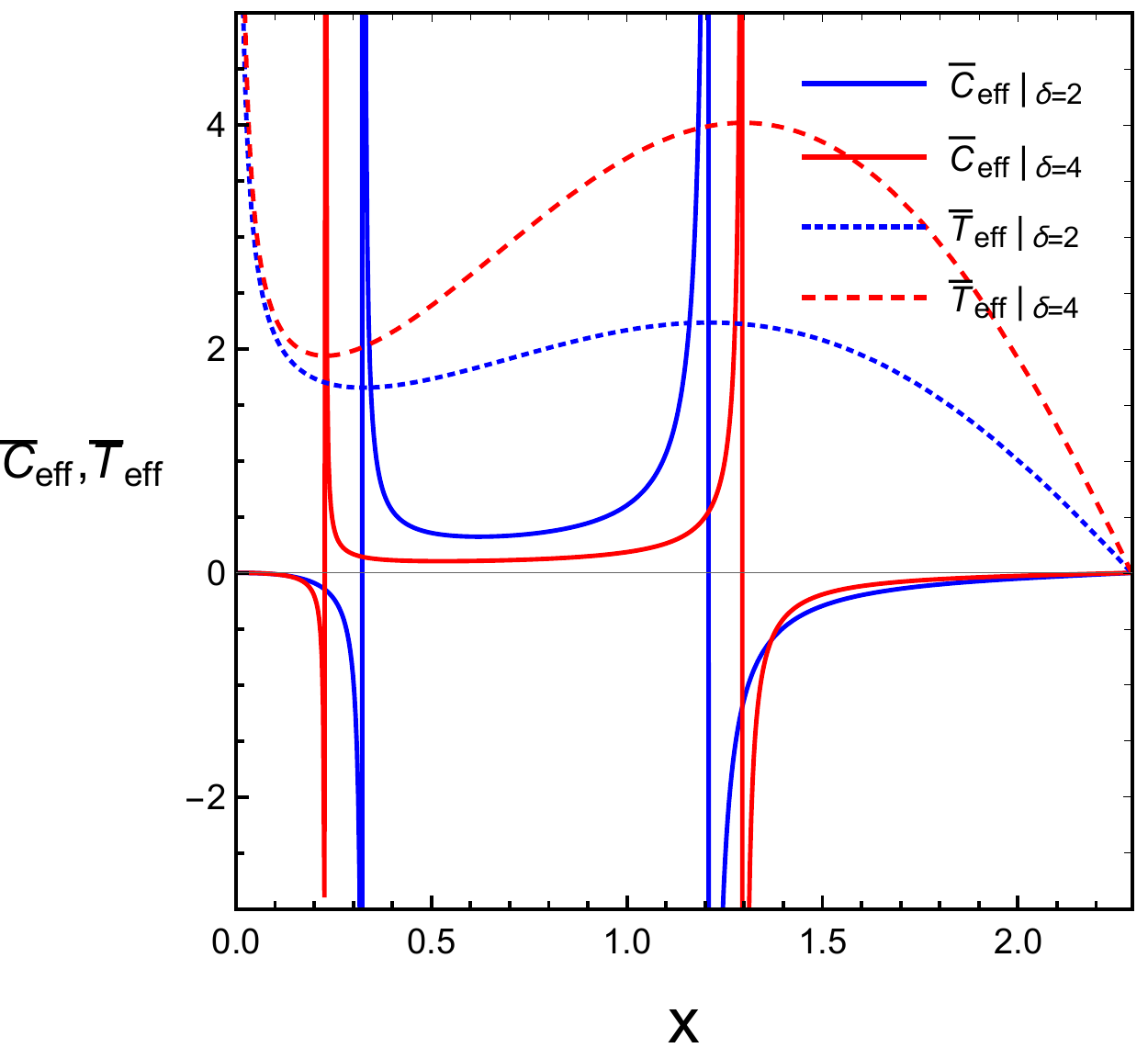}
	\end{center}
	\caption{Left panel shows the effective heat capacity profile with various values of $\delta$ by fixing $b_0=0.1$, $b_1=0.7$. Right panel shows the effective heat capacity profile compare to the temperature profile by fixing $b_0=0.1$, $b_1=0.7$.}\label{fig:Ceff}
\end{figure}

Now, let us consider the heat capacity at constant pressure. The thermodynamic system is locally stable if  heat capacity is positive. The thermodynamic system with negative heat capacity will radiate thermal energy. Then, the system gets hotter and will be lost more thermal energy via the radiation. In other words, the hotter the black hole is, the more it radiates, and then the system will vanish eventually. The heat capacity of the effective system can be found by
\begin{align}
	C_\text{eff}&=\Big(\frac{\partial M}{\partial T_\text{eff}}\Big)_P=\frac{\left(\partial_{r_b} M \right)_P}{F(r_b,r_c)}.
\end{align}
Note that $M$ can be rearranged to express in terms of only $r_b$, so that we can take derivative with respect to $r_b$ while the pressure is kept to be constant. Note also that the explicit expression of effective heat capacity is very lengthy, we do not show here. The important key is that the heat capacity diverges at the extrema of the temperature and the positive part corresponds to the positive slope of the temperature as found in the right panel in Fig. \ref{fig:Ceff}. Note that we have used the dimensionless variables $\bar{C}_\text{eff}=\frac{C_\text{eff}}{\pi 2^{2/3}r_V^2}$ for the plot in Fig. \ref{fig:Ceff} and the variable $r_c$ is transformed to $r_b$ by using relation in Eq. \eqref{y+}. From the left panel in this figure, one can see that the heat capacity is always negative for $\delta < \delta_{\text{eff}}$ and the middle part becomes positive for $\delta > \delta_{\text{eff}}$. This is compatible to the analysis of the temperature profile. Summarily, for the effective system approach, the moderated-sized black string is locally stable for $\delta > \delta_{\text{eff}}$. 

\begin{figure}[h!]
	\begin{center}
		\includegraphics[scale=0.6]{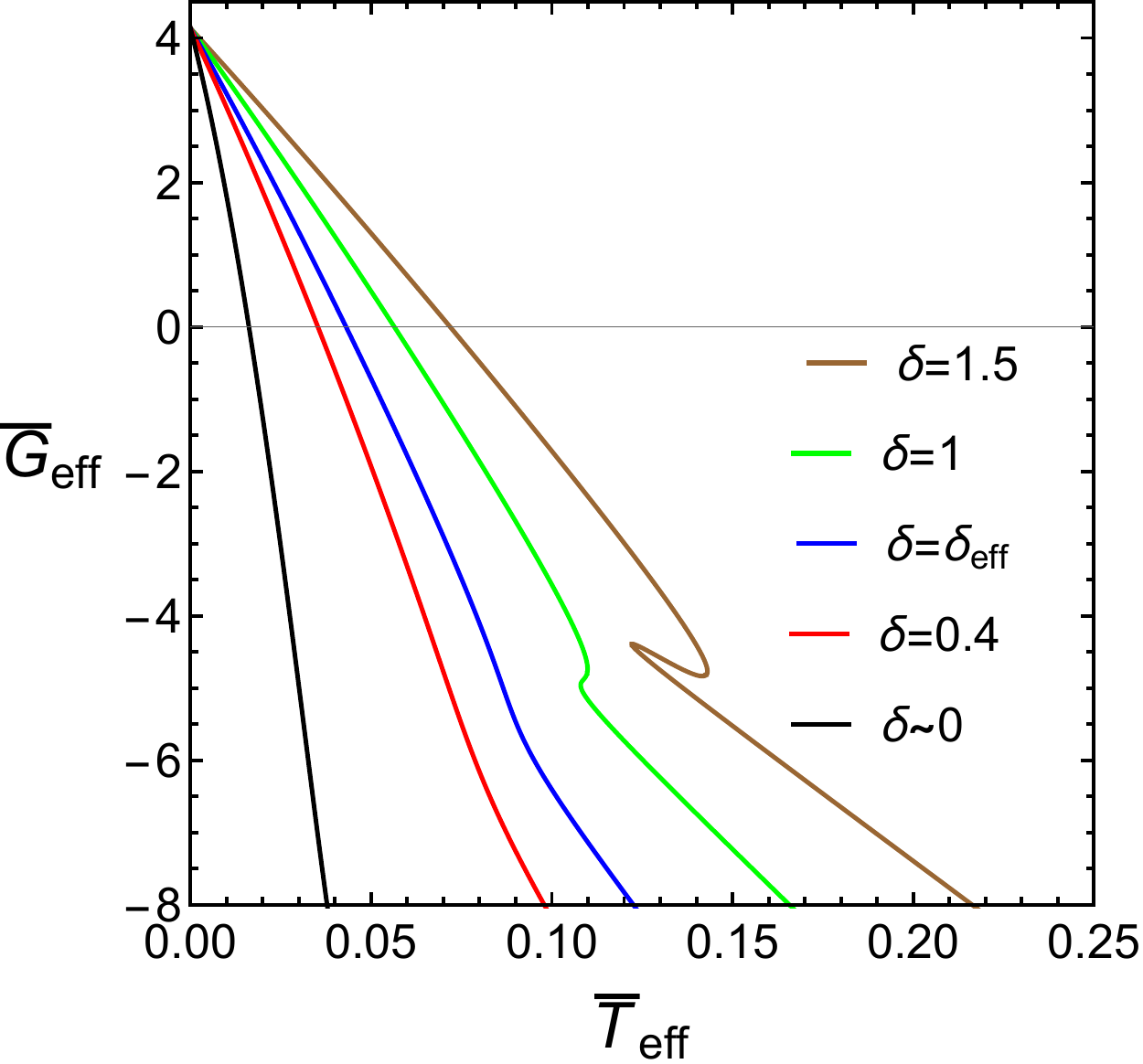}\quad
		\includegraphics[scale=0.285]{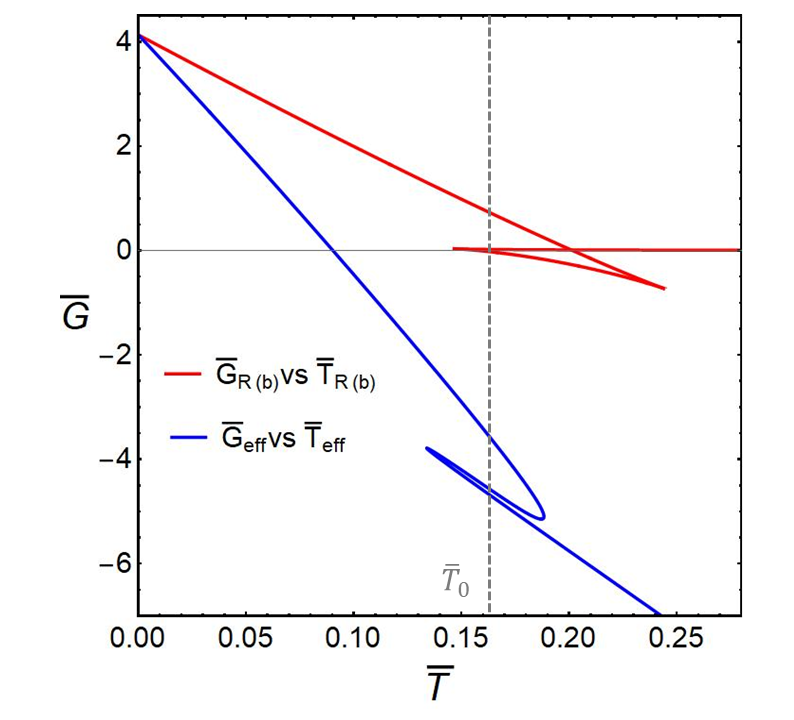}
	\end{center}
	\caption{Left panel shows the profile of the dimensionless Gibbs free energy $\bar{G}_\text{eff}=\frac{2^{4/3}}{M}G_\text{eff}$ against the effective temperature with various values of $\delta$ by fixing $b_0 =0.1, b_1=0.7$. The right panel  shows the comparison of the Gibbs free energy for separated system approach and effective system approach with $b_0 =0.1, b_1=0.7$ and $\delta=2.15$.}\label{fig:GeffT}
\end{figure}

Let us move to consider the global stability, which can be determined by considering the value of the Gibbs free energy. According to effective description, the effective Gibbs free energy can be written as
\begin{align}
	G_\text{eff}&=M-T_\text{eff}\Big(S_{R(b)}+S_{R(c)}\Big),\label{Geff}
\end{align}
The behavior of the effective Gibbs free energy is shown in the left panel in Fig. \ref{fig:GeffT}. From this figure, it is found that the Gibbs free energy corresponding to the locally stable size of the black string is always negative. Therefore, the locally stable black string is always globally stable. This issue is different from one for the separated system approach in which there is a part of parameter space corresponding to the positive value of Gibbs free energy. Hence, there is no other bound of $\delta$ for the Gibbs free energy in effective description. Note that the part between cusps of $\bar{G}_\text{eff}$ corresponds to the moderate-sized black string. It is important to note also that the slope of the lines plotted in the left panel in Fig. \ref{fig:GeffT} does not correspond to the entropy, since the negative sign is needed to add in the change in entropy at the cosmic horizon. This is the consequence from the definition of the effective quantities as we discussed previously. As a result, the cusps for the effective system approach is not sharp as found in one for the separated system approach. This is also see from the right panel in Fig. \ref{fig:GeffT}.

Moreover, it is found that the effective free energy at a certain temperature is more negative compared to one for the system evaluted at the black string horizon as shown in the right panel in Fig. \ref{fig:GeffT}. By comparing to the free energy of the hot gas which is zero, it is found that the Gibbs free energy is discontinuous. As a result, for the effective system approach, the hot gas phase has to undergo a zeroth-order phase transition in order to evolve into the moderate-sized stable black string in the effective system. This is one of the key differences to the separated system approach in which the transition is the first-order type as we have discussed.

\section{Conclusion}\label{sec: concl}
The black string is a result from cylindrically symmetric solution of Einstein field equation with cosmological constant analogous to the black hole which is a result from spherically symmetric solution. With respect to the cylindrical symmetry, an event horizon is a cylindrical shell of radius $r_h$ analogous to the Schwarzschild radius of the black hole. It is found that the thermodynamic description of the black string is in the same form as one for the black hole. For example, the entropy of the black string is proportional to its area and then the equivalent laws of thermodynamics are in the same form. The thermodynamic properties of the black string have been investigated. Recently, the black string solutions in modified gravity are investigated and their thermodynamic properties have also been studied. In the present work, we aim to investigate the thermodynamic properties of the black string in dRGT massive gravity which is one of viable models of modified gravity theories. One of the important signatures of the black string solution is that it is an asymptotically dS/AdS spacetime. Therefore, the investigation can be classified as asymptotically dS solution and asymptotically AdS solution. For asymptotically AdS solution, it has been investigated \cite{Tannukij:2017jtn,Ghosh:2019eoo,Hendi:2020apr}, while it is very rare for the asymptotically dS solution. This may come from two main obstructions. Firstly, there are two horizons and then they correspond to two thermal systems with generically different temperatures. Therefore, the systems are out of thermal equilibrium and then the thermodynamic variables defined in equilibrium state cannot be properly used. Secondly, the system is locally unstable. In this paper, we consider the effective system to overcome the first obstruction and using R\'enyi entropy to investigate the possibility to obtain the stable dRGT black string.  

We begin the investigation by showing that the existence of two horizons is a generic property of the dRGT black string with asymptotically dS spacetime. Moreover, we show that it is not possible to have horizons for the Lemos's black string with asymptotically dS spacetime. For the model parameters of massive gravity theory $(b_1, b_0)$, the region for existence of the horizons is shown explicitly in the left in Fig. \ref{fig:horizon}. From the existence of two horizons, we divide our consideration into two parts, separated system approach and effective system approach. 

For the separated system approach, the thermodynamic systems can be investigated separately by assuming that the systems are far enough and the temperatures of the systems are not significantly different. By adopting the first law of the black string, we examined the nonextensivity by replacing the R\'enyi entropy with the GB ones. As a result, we obtained the empirical temperature defined in the same way as performed in the GB case. By analyzing the slope of the temperature, we found that it is possible to obtain the locally stable black string. The lower bound on the nonextensive parameter is found as $\delta_c = 3/(2^{10/3}b_1^2)$. We further investigate the global stability by considering the sign of the Gibbs free energy. By requiring that the Gibbs free energy must be negative, it is found that the lower bound is stronger than one obtained from local stability, $\delta_G > \delta_c$. This implies that in order to form the black string the nonextensive nature of the black string must be large enough. Note that, in the viable range of the black string system $\delta > \delta_G$, the system evaluated at the cosmic horizon is both locally and globally stable. Furthermore, we found that it is possible to obtain the first-order Hawking-Page phase transition which is the transition between the thermal radiation or hot gas phase and the stable black string phase. This is one of important results to distinguish between two approaches, since it is the first-order phase transition for the separated system approach. At this stage, one can roughly argue that the structure of the graviton mass provides the existence of the asymptotically dS spacetime of the dRGT black string and the nonextensivity provides the stability of the black string.    

For the effective system approach, we still restrict on the first law which is in the same form with one in separated system approach. However, the temperature and volume are treated as the effective quantities where the total entropy is in the additive form of those of the separated systems. The effective quantities are defined by using the criterion such that the heat flow of the system evaluated at the cosmic horizon has the opposite sign to one at black string horizon \cite{Nakarachinda:2021jxd}. This comes from the fact that the observer stays between the black string horizon and cosmic horizon. It is interesting that other thermodynamic variables will relate to the effective quantities in the same way as in the GB statistics. As a result, the heat capacity and the free energy can be well-defined and thermodynamic stability can be properly investigated. By considering the local stability, it is possible to find the region of parameter space with positive heat capacity. This can be performed in the same way as done in separated system approach by considering the slope of the temperature. As a result, the moderate-sized black string is found to be locally stable with large enough value of the nonextensivity. The lower bound of the nonextensivity can be obtained by analyzing the possibility of existence of the local extrema of the temperature. We found that the bound in effective system approach is stronger than one in separated system approach, $\delta_{\text{eff}} =1.254 \delta_c$. This implies that the thermodynamic stability of the black string in the effective system approach requires nonextensive nature of the system greater than one in the separated system approach. Moreover, we also found the way to distinguish the black string from both approaches. In particular, we found that there exist particular temperatures in which the black string in both approaches will be locally stable. In this case, the black string in the effective system approach is always larger than one in the separated system approach. Moreover, there exist particular temperatures for which only black string in the effective or separated system approach is stable. As a result, these particular temperatures can be used to distinguish between the two approaches. For the global stability, we found that the Gibbs free energy in the range with local stability is always negative. Therefore, the locally stable black string is always globally stable without another requirement as found in separated system approach. Moreover, the Hawking-Pages phase transition is found to undergo from hot gas to the black string with the zeroth-order phase transition. It is worthwhile to argue that it is suitable to use the notion of the effective system defined in the present work, since the singularities in the effective quantities can be avoided and the main equations in thermodynamics can be systematically obtained. Moreover, it is interesting to apply this notion of the effective system to investigate thermodynamic properties of other black holes/strings with multiple horizons especially dRGT black hole. We leave this investigation for further works.  

For the dRGT black string, the first law of thermodynamics can be formulated by interpreting the massive gravity parameters $c_1$ and $c_0$ as dynamical variables. In the present work, we restrict on the system with fixing $c_1$ and $c_0$. Therefore, it is worthwhile to extend the results in the present paper to more general one by keeping  $c_1$ and $c_0$ as dynamical variables. In this case, the physical interpretation of the new variables should be carefully discussed and the new criterion for defining the effective quantities must be introduced. Moreover, it is possible to consider more general black string solutions such as charged and rotating black string solutions. Therefore, the mentioned solutions may provide more interesting phenomena such as the critical behavior of the thermodynamic system. These are beyond our scope of the present paper and supposed to investigate in further works.

It is important to note that the first law of the thermodynamics for the black string in R\'enyi statistics is not directly related to one in geometrical description as found in GB statistics. In particular, we adopt the first law in Eq. \eqref{1stlaw-Renyi} by using the similar form with one in GB statistic. This allows us to investigate the nonextensive nature of the black string based on the R\'enyi statistics while the structure of the thermodynamics is still in the same form with one in GB statistics. Therefore, the R\'enyi temperature cannot be directly derived from the description of quantum field theory to curved spacetime and then it may not be interpreted as physical quantity as discussed in Ref. \cite{Nojiri:2021czz}. However, in the derivation of the Hawking temperature in Ref. \cite{Hawking:1976ra} seems to be based on the GB statistics corresponding to the extensive system while the resulting entropy is nonextensive. Therefore, while the entropic nature of a nonextensive system can be debatable, there is no guarantee that the Hawking temperature of a black hole/string must be derived using the GB statistics.

One of worthy features of the R\'enyi entropy is that it relates to the entanglement entropy in certain limit as discussed in Ref. \cite{Promsiri:2021hhv}. In particular, for large nonextensivity limit, the R\'enyi entropy can be expressed in the form $S_R \sim \ln(r_h/L_\lambda)$ which coincides with one for the entanglement entropy $S_e \sim \ln(\xi/a)$ \cite{Vidal:2002rm,Calabrese:2004eu}. Note that $r_h, L_\lambda, \xi$ and $a$ denote horizon radius, nonextensive length, correlation length and lattice spacing, respectively. Note also that the black string radius and nonextensive length are scaled by $r_h \sim b_1 r_V$ and $L_\lambda \sim \frac{1}{\sqrt{\delta}}r_V$. Therefore, there exists the critical value obtained from the lower bound of nonextensive parameter such as $(r_h/L_\lambda)_c = \sqrt{\delta_c b_1^2} = \sqrt{3/(8\times2^{1/3})}$ for the local stability in separated system approach. In this sense, the stable black string can be formed by requiring that its radius must be greater than the critical value $r_h > \sqrt{\delta_c b_1^2} L_\lambda $ while, in the limit $r_h \gg \sqrt{\delta_c b_1^2} L_\lambda $, the R\'enyi entropy will take the same form with entanglement entropy. In gravitation point of view, for the length scale which is much larger than the Vainshtien radius $r\gg r_V$, the graviton mass plays a major part of the gravitational interaction while it is suppressed at the scale below the Vainshtien radius $r_{\Lambda_3} \ll r\ll r_V$. Note that $r_{\Lambda_3}$ is the cutoff scale of the dRGT massive gravity theory. Below this length scale, the theory is supposed to be not trustable as discussed in Sec. \ref{sect:dRGT BS}. Combining this notion with the limit to obtain the entanglement entropy form, we may argue that the black string radius can play the role of the correlation length and then the gravitational interaction contributed from the graviton mass can play the role long-range interaction inherited from the entanglement description for the limit $r_h \gg r_V$.  This may shed light on the interplay between nature of quantum entanglement and gravitational interaction contributed from the graviton mass. 

\section*{Acknowledgement}
We are grateful to Lunchakorn Tannukij, Ekapong Hirunsirisawat, Chatchai Promsiri  and Isara Chantesana for helpful discussion. This research project is supported by National Research Council of Thailand (NRCT): NRCT5-RGJ63009-110 and SERB-DST, India for the ASEAN project IMRC/AISTDF/CRD/2018/000042. PS is supported by The Prof. Dr. Sujin Jinahyon Foundation. PW is supported by National Science, Research and Innovation Fund (SRF) through grant no. P2565B202.

\end{document}